\documentclass[a4paper,12pt]{fullarticle}
\usepackage[british]{babel}
\usepackage{csquotes}
\usepackage{sciencestuff}
\usepackage[algoruled,algosection,vlined,shortend,linesnumbered]{algorithm2e}

\usepackage[dvipsnames]{xcolor}
\usepackage{esint}
\usepackage{booktabs}
\usepackage{array}
\usepackage[table]{xcolor}
\usepackage{graphicx}
\usepackage{xcolor}
\usepackage{nicematrix}
\usepackage{booktabs}
\usepackage{listings}
\usepackage{mdframed}
\usepackage{tcolorbox}
\usepackage{hyperref}

\usepackage{multicol}
\title{%
  Relational background field formulation of discrete quantum gravity models: \\ \relax
    {\Large The case of $O(N)$ vector models in zero dimensions}}

\author[1]{Vincent Lahoche\emailfoot{vincent.lahoche@cea.fr}}
\author[2]{Dine Ousmane Samary\emailfoot{dine.ousmanesamary@uac.bj}}
\author[1]{Parham Radpay\emailfoot{parham.radpay@cea.fr}}

\affil[1]{%
  Université Paris-Saclay, CEA, %LIST,
  \protect \\
  Palaiseau, F-91120, France
}
\affil[2]{%
	Faculté des Sciences et Techniques (ICMPA-UNESCO Chair)
	\protect \\
	Université d'Abomey-Calavi, 072 BP 50, Benin
}

\date{}

\hypersetup{%
  pdftitle={RG-O(N)-vector},
  pdfkeywords={%
  functional renormalization group,
  theoretical physics,
  data science,
  signal analysis,
  signal detection,
    random matrix theory
    },
  pdfsubject={signal detection}
}

\newtheorem{remark}{Remark}
\newtheorem{definition}{Definition}
\newtheorem{theorem}{Theorem}

\newtheorem{claim}{Claim}

\begin{document}

\maketitle

\begin{abstract}
In a recent paper \href{https://arxiv.org/abs/2609.21031}{https://arxiv.org/abs/2609.21031}, we introduced the concept of a relational background field as a new approach to the renormalization group (RG) for discrete quantum gravity models, focusing on random matrix models. The proposed method offers the dual advantage of preserving the model's gauge symmetry (typically $U(N)$ or $O(N)$), whose breaking in conventional approaches has remained an open problem, and providing a more satisfactory treatment of the continuum or infrared (IR) limit. In this limit, the theory is described by a non-local field theory in dimension $D = 2k + 3$, where $k$ denotes the multicriticality index at the spectral edge of the effective spectrum as dictated by random matrix theory. This pedagogical paper reviews this construction for real random vector models in zero dimensions, with $O(N)$ as the symmetry group.
Following the general strategy of background-field-type formulations, we show, using the BBP (Baik–Ben Arous–Péché) phase transition, that a partial Hubbard–Stratonovich decomposition of the quartic interaction, involving a matrix-valued intermediate field, allows a preferred notion of scale to emerge. This scale, in turn, defines an RG flow, whose perturbative and non-perturbative structure is analyzed. Our calculation demonstrates the existence of an \textit{asymptotic} Wilson–Fisher-type fixed point with a single relevant direction, and we discuss its critical exponent and physical interpretation, also in relation to the double-scaling limit of these models. 
\end{abstract}

\highlights{%
We exploit the self-averaging property of large random matrices to construct a novel renormalization group scheme that is "safe" with respect to the underlying gauge symmetry, based on the effective spectrum of the intermediate Hubbard–Stratonovich field. This method provides a new notion of scaling for discrete quantum gravity models.
}


\clearpage
\tableofcontents
%{\small\tableofcontents}

\clearpage

%%%%%%%%%%%%%%%%%%%%%%%%%%%%%%%%%%%%%%%%%%%%%%%
%                                             %
% Introduction                                %
%                                             %
%%%%%%%%%%%%%%%%%%%%%%%%%%%%%%%%%%%%%%%%%%%%%%%

\section{Introduction}\label{introduction}

The last ten years have seen significant progress in the field of group field theories (GFTs), a covariant formulation of loop quantum gravity (LQG) whose perturbative expansions generate spin foam amplitudes \cite{freidel2005group,rovelli2004quantum}. These advances mainly concern the realization of the idea that spacetime can emerge from the collective behavior of a large number of pre-geometric quantum degrees of freedom, the quanta of space, within a hydrodynamic limit. In standard approaches, this hydrodynamic phase takes the form of a condensate, one of whose main features is to reproduce, at the mean-field level, a non-singular Friedmann cosmological dynamics—that is, one free from the Big-Bang singularity, see \cite{calcinari2026collectiveexcitationsquantumgravity,oriti2017universe,oriti2017bouncing,oriti2016emergent,oriti2016horizon,gielen2016quantum} and references therein. The existence of these condensates is physically motivated, within the mean-field approximation, by the fact that the Heisenberg dynamics of free GFTs naturally leads to the macroscopic occupation of a single quantum state. Furthermore, mean-field theory itself is motivated by a special property of GFTs in hyperbolic geometry, showing, under certain assumptions, that the \textit{Ginzburg criterion} is satisfied in a condensed phase \cite{Marchetti_2023,Marchetti2_2023,Dekhil_2025}. In other words, this result tells us that if the system is in the condensed phase, the Gaussian fluctuations around this specific vacuum are not strong enough to destroy it. However, although this result is of great value in demonstrating the consistency of the condensation hypothesis, it does not prove the existence of a renormalization group (RG) flow connecting the symmetric phase in the ultraviolet (UV) to this condensed phase in the IR. This latter point remains a challenge today and would be necessary to definitively substantiate the geometrogenesis scenario. Finally, this criterion ignores certain subtleties specific to GFT non-localities, notably the role of the anomalous dimension \cite{lahoche2020pedagogical,lahoche2018nonperturbative}.

The construction of a reliable RG for GFTs has been the subject of extensive work over the past fifteen years, whether through formal results such as proofs of BPHZ-type renormalizability theorems \cite{carrozza2014tensorial,carrozza2014renormalization,carrozza2016flowing,geloun2013renormalizable,lahoche2015renormalization}, proofs of the constructibility of these models \cite{Rivasseau_2021}, or non-perturbative approximations \cite{lahoche2020pedagogical,lahoche2018nonperturbative,lahoche2017functional,benedetti2015functional,benedetti2016functional,lahoche2019ward,geloun2016functional,Ben_Geloun_2024,yerima2026functionalrenormalizationgrouprank4,lahoche2025stochasticdynamicsgroupfield,Lahochesto_2023,Pithis_2020}. The latter have revealed a significant portion of the difficulties specific to GFTs, related in particular to non-trivial Ward identities inherent in the choice of the kinetic term, especially in so-called "tensorial" models, where the symmetry group structuring the interactions is explicitly broken by a "Kontsevich-like" kinetic kernel \cite{gurau2024quantum}. The choice of this symmetry-breaking kinetic term is all the more essential as the hydrodynamic limit discussed previously heavily depends on it. Notably, in \cite{lahoche2020pedagogical,lahoche2018nonperturbative}, the authors show how modified Ward identities cause the non-perturbative fixed points that seemed natural in tensorial GFTs without a closure constraint to disappear \cite{Lahoche_2021}.

In parallel, the RG has also been considered as an approach to continuous phase transitions in random tensor models \cite{gurau2017random}, whose interaction structure inspired tensorial GFTs \cite{carrozza2014tensorial,Rivasseau_2013,Rivasseau_2014} and for which, unlike random matrix models for 2D gravity, no complete analytical solution is available. Current approaches are essentially non-perturbative versions \cite{eichhorn2013continuum,eichhorn2018flowing,eichhorn2019towards,castro2026quantitativecharacterizationgravitationaluniversality,Eichhorn_2020,Eichhorn2_2020} of a framework initially proposed by Zinn-Justin and Brézin \cite{brezin1992renormalization} for matrix models, allowing for the qualitative reproduction of the critical exponents of the double-scaling limit. This method relies on a Wilsonian approach in which a subset of matrix or tensor degrees of freedom is partially integrated out, thereby generating a flow for the coupling constants. However, this method poses a number of problems that jeopardize its reliability. The first problem stems from the breaking of the symmetry structuring the interactions (typically $U(N)$ or $O(N)$) required by the partial integration procedure. Indeed, in ordinary field theories, there is a preferred ordering dictated by the size of the Gaussian fluctuations (the spectrum of the free propagator). But for a strictly $U(N)$-invariant matrix or tensor model, the kinetic term is always proportional to the identity, and no privileged ordering exists: consequently, any choice necessarily breaks the structure group. This breaking has hard-to-control consequences, including the proliferation of interactions that also violate this symmetry, as dictated by the modifications of the Ward identities required by the regularization procedure \cite{Lahoche:2020pjo,Lahoche:2019ocf}. Another problem stems from the fact that this approach, in some sense, pits the continuum limit against the infrared limit ($N\to 0$), which notably appears poorly defined and inconsistent with the approximations used to compute the flow equations \cite{gurau2024quantum}. Note that the problems inherent to this RG approach are not disconnected from the previous issue concerning GFTs; these models share a common interaction structure with the latter, so much so that most of the difficulties encountered in constructing a reliable RG for these models also apply to GFTs.

In a recent work \cite{Lahoche_2026_background}, we showed that, for Hermitian random matrices, another path is possible, inspired by "background field" type approaches \cite{reuter1998nonperturbative}. In these approaches, the fluctuations of the gravitational field $g_{\mu\nu}$ in the typical Euclidean "matter-field" partition function, involving, for instance, many species of fermionic matter fields and bosonic gauge fields:
\begin{equation}
Z=\int \dd g_{\mu\nu} \int \prod_i \dd \psi_i \dd \bar{\psi}_i \prod_j \int \dd A_j \, e^{-\mathcal{S}(g_{\mu\nu},\psi,\bar{\psi},A)}\,,
\end{equation}
are split into two components, $g_{\mu\nu}=\bar{g}_{\mu\nu}+h_{\mu\nu}$, a background field $\bar{g}_{\mu\nu}$ and a "fluctuating" field $h_{\mu\nu}$,
\begin{equation}
Z=\int \dd \bar{g}_{\mu\nu} \int \dd h_{\mu\nu} \int\prod_i \dd \psi_i \dd \bar{\psi}_i\prod_j \int \dd A_j \, e^{-\mathcal{S}(g_{\mu\nu},\psi,\bar{\psi},A)}\,.
\end{equation}
Most applications then focus on a regime where the fluctuations of the background field become negligible, so that it essentially freezes onto a classical configuration $\bar{g}_{\mu\nu}^{(0)}$:
\begin{equation}
\int \dd h_{\mu\nu} \int \prod_i \dd \psi_i \dd \bar{\psi}_i \prod_j \int \dd A_j \, e^{-\bar{\mathcal{S}}(\bar{g}_{\mu\nu}^{(0)},h_{\mu\nu},\psi,\bar{\psi},A)}\,.
\end{equation}
In this new action $\bar{\mathcal{S}}$, the field $\bar{g}_{\mu\nu}^{(0)}$ ceases to be a dynamical entity and, by defining the Laplace or Dirac operators, thereby defines the notion of scale from which it is possible to construct an RG for the remaining gauge and matter fields $({A_j},{\psi_i},{\bar{\psi}_i})$.

We have shown that, in the case of random matrices, it is also possible to isolate a subset of the degrees of freedom which, in the continuum limit, "freeze" into a background essentially decoupled from the fluctuations, thereby defining an intrinsic notion of scale for the latter. This approach has the undeniable advantage of explicitly preserving the structural symmetry ($U(N)$ for the model considered in the paper) and of possessing a well-defined IR limit in the continuum limit $N\to \infty$, so that the planar approximation for the flow equations is always justified. The method is obviously not limited to random matrices, and its application to tensors will be the subject of future work. In this pedagogical note, we present the method for real random vectors with $O(N)$ structural symmetry, where it remains technically simpler while staying structurally very close to its formulation for tensors, which share, in the melonic sector, a common structure of branched polymers.

\begin{remark}
The zero-dimensional $O(N)$ model is completely solvable by elementary means, and the main purpose of this paper is to serve as a pedagogical and analytical benchmark for the relational background field formalism. In this setting, the large-$N$ flow remains technically manageable and admits an exact non-perturbative solution within the Local Potential Approximation, allowing us to test the construction directly against known exact results, clarify the physical meaning of the perturbative IR fixed point, and identify the mechanism through which the non-confining criticality and double-scaling behavior are recovered. 
\end{remark}

\paragraph{Outline.} The outline of the paper is as follows. In Section \ref{summary}, we begin by recalling the definitions and certain elementary properties of O(N) vector models, in particular regarding their double-scaling limit, and the construction of the Zinn-Justin and Brézin Wilsonian RG. In Section \ref{relationalback}, we present the relational background field method, and we show in particular how a notion of scale can emerge in the continuum limit without breaking the native O(N) symmetry of the model. Furthermore, this notion of scale is now subordinated to the continuum limit, which guarantees the "freezing" of this reference background. In Section \ref{renormalization}, we show perturbatively how this emergent scale allows for the definition of an RG. Section \ref{nonperturbative} in turn investigates the non-perturbative regime; and Section \ref{conclusion} provides an extended discussion and perspectives for future work.

\section{A digest of the $O(N)$ vector model in zero dimensions}\label{summary}
In this section, we review the general definition of vector $O(N)$ models and discuss some of their main properties, with particular emphasis on their large-$N$ analysis and multi-scaling behavior. We conclude with a review of an RG approach proposed by Zinn-Justin, which allows for a qualitative recovery of the multi-scaling critical exponents. For a general review of these topics, the reader is referred to \cite{Moshe_2003,Zinn_Justin_2014}. Readers who are already familiar with this subject may simply skim through this section.

\paragraph{Definition of the model.} The $O(N)$ vector model is usually defined by an integral of the form:
\begin{equation}
Z_N:=\left(\frac{N}{2\pi}\right)^{N/2}\int \, \dd \phi\, e^{-N V(\phi^2)}\,,
\end{equation}
where $\phi \equiv (\phi_1, \phi_2, \cdots, \phi_N) \in \mathbb{R}^N$, and $V(x)$ is a polynomial whose normalization is chosen such that\footnote{A less restrictive condition is to assume that $V(x)$ is analytic in the neighborhood of the positive real axis.} $V(x) = x/2 + \mathcal{O}(x^2)$. Finally, the prefactor in front of the integral normalizes the Gaussian measure. Equivalently, the model can be defined by specifying the probability density $P[\phi]$ for a given configuration $\phi$:
\begin{equation}
P[\phi]:=\left(\frac{N}{2\pi}\right)^{N/2}\,\frac{e^{-N V(\phi^2)}}{Z_N}\,.\label{probaphi}
\end{equation}
The partition function can be evaluated using the saddle-point method in the limit $N\to \infty$. Passing to angular variables $\phi=(\sqrt{\rho},\Omega)$, and exploiting the rotational invariance of the potential:
\begin{equation}
Z_N=\left(\frac{N}{2\pi}\right)^{N/2}\int_0^\infty \rho^{\frac{N-1}{2}} \dd \sqrt{\rho} \int_\Omega \dd \Omega\, e^{-N V(\rho)}\,,
\end{equation}
by isolating the angular integral, we find that, with $\int_\Omega \mathrm{d}\Omega := \frac{2\pi^{N/2}}{\Gamma(N/2)}$,
\begin{equation}
Z_N=\mathcal{C}(N)\int_0^\infty \frac{\dd \rho}{\rho} \, e^{-N W(\rho)}\,,
\end{equation}
where, at large $N$, using Stirling's formula, $\mathcal{C}(N)\approx e^{N/2}\sqrt{N/4\pi}$, and:
\begin{equation}
W(\rho):=V(\rho)-\frac{1}{2} \ln \rho\,.
\end{equation}
The saddle point $\rho_0$ is given by the condition that the first derivative $W^\prime(\rho)$ vanishes, which immediately yields:
\begin{equation}
2V^\prime(\rho_0)\rho_0=1\,,
\end{equation}
For the quartic model, $V(x)=\frac{1}{2}x+\frac{g}{4} x^2$, we have in particular:
\begin{equation}
g \rho_0^2+\rho_0-1=0\,.\label{condition_sadle_quart}
\end{equation}
This equation admits two solutions, of which we retain only the one tending to $1$ in the limit $g\to 0$:
\begin{equation}
\rho_0=\frac{-1+\sqrt{1+4g}}{2g}\,.\label{solutionrho0}
\end{equation}

\begin{remark}
The integral can be evaluated in terms of hypergeometric functions for any $N$. For the quartic model in particular, using the definition of Weber's parabolic cylinder function $D_p(z)$:
\begin{equation}
\int_0^\infty dx \, x^{\nu-1} e^{-\beta x - \gamma x^2} = (2\gamma)^{-\frac{\nu}{2}} \Gamma(\nu) e^{\frac{\beta^2}{8\gamma}} D_{-\nu}\left(\frac{\beta}{\sqrt{2\gamma}}\right)\,,
\end{equation}
which are furthermore related to confluent hypergeometric functions of the first kind.
\end{remark}

\paragraph{Multicritical points, double scaling.} The saddle point identified previously is a special case of a multicritical point, the general definition of which is given here:
\begin{definition}
A multicritical point of order $m$ ($m\geq 2$) is a point $\rho_0$ where the first $(m-1)$ derivatives of $W(\rho)$ vanish, while $W^{(m)}(\rho_0)\neq 0$:
\begin{equation}
W^\prime(\rho_0)=\cdots=W^{(m-1)}(\rho_0)=0\,,\qquad W(\rho)-W(\rho_0) \propto (\rho-\rho_0)^m+\mathcal{O}((\rho-\rho_0)^{m+1})\,.
\end{equation}
\end{definition}
The value $m=2$ corresponds to the previous saddle point, whose fluctuations are Gaussian. For $m>2$, the fluctuations are no longer Gaussian, and the integration contour must be deformed in the complex plane along the principal directions of the critical point, given by the condition $\Re((\rho-\rho_0)^m)>0$.

Reaching a multicritical point of order $m$ requires fixing $(m-1)$ parameters — the first $(m-1)$ derivatives of $W(\rho)$:
\begin{equation}
W^\prime(\rho_0)=W^{\prime\prime}(\rho_0)=\cdots=W^{(m-1)}(\rho_0)=0\,.
\end{equation}
Up to a change of variables, the path integral can be written at the critical point as:
\begin{equation}
Z_N \sim \int \dd z \, e^{-N z^m}
\end{equation}
This integral fixes the typical scale of the fluctuations $z \sim N^{-1/m}$, and shows that the $(m-1)$ parameters tuned to vanish the derivatives are relevant from a power-counting perspective. Indeed, the rescaling $z = N^{-1/m} s$ shows that a perturbation $\delta V = a N z^n$ scales as $N^{1-n/m}$, so that perturbations with $n > m$ are irrelevant. A perturbation at the critical point will thus take the general form, up to irrelevant terms:
\begin{equation}
\delta V = \sum_{n=1}^{m-2} N^{1-n/m} a_n \, s^n\,.
\end{equation}
Note that the $(m-1)$-th order term, which should in principle appear in the general expression, is absent here, since it can be directly generated by the dominant term $s^m$ through a suitable constant translation of the parameter $s$. The scaling transformation highlights power counting and shows that the model is characterized by the values of the couplings $b_n := N^{1-n/m} a_n = \mathcal{O}(1)$, with all models having the same values for these couplings sharing the same $N \to \infty$ limit. In other words, the multicritical {scaling} is characterized by the fact that the couplings $a_n$ vanish as $N^{n/m-1}$.

Let us analyze the double-scaling limit in more detail for a quartic model, corresponding to the case $m=3$. In this case, $\delta V = N^{2/3} a_1 s$, and the integral reads:
\begin{equation}
Z_N \sim \int \dd s \, e^{-s^3-N^{2/3} a_1 s}\,.\label{Z_Ndoublescal}
\end{equation}
The criticality condition for $\rho_0$, $W^{\prime\prime}(\rho_0)=0$, reads explicitly here:
\begin{equation}
V^{\prime\prime}+\frac{1}{2} \frac{1}{\rho_0^2}=0\quad \to \quad g \rho_0^2=-1\,.\label{critical_cond}
\end{equation}
By combining this condition with condition \eqref{condition_sadle_quart}, we obtain $\rho_0=2$, which, together with \eqref{critical_cond}, yields $g_c=-1/4$. Note that this critical value of $g_c$ is also the boundary of the domain of solution \eqref{solutionrho0}, below which $\rho_0$ is imaginary. In the vicinity of $g_c$,
\begin{equation}
\rho_0(g)-\rho_0(g_c)\sim (g-g_c)^{1/2}\,.
\end{equation}

An alternative and instructive perspective on double scaling is obtained by studying the behavior of the free energy $F_N(g) := -N^{-1} \ln Z_N$ (up to an additive $g$-independent constant coming from $\mathcal{C}(N)$). At the saddle point, 
\begin{equation}
F_N(g) \approx  W(\rho_0)=\frac{1}{2} \rho_0+\frac{g}{4}\rho_0^2{-\frac{1}{2}\ln \rho_0}\,.
\end{equation}
Because $\partial_\rho W(\rho_0)=0$, we have:
\begin{equation}
\frac{\dd}{\dd g} F_N= \frac{\dd \rho_0}{\dd g}\frac{\partial}{\partial \rho_0} W(\rho_0(g))+\frac{\partial}{\partial g} W(\rho_0(g))=\frac{1}{4}\rho_0^2\,,
\end{equation}
and therefore:
\begin{equation}
\frac{\dd}{\dd g} F_N\Big\vert_{\text{sing}} \sim (g-g_c)^{1/2} \quad \to \quad F_N \sim (g-g_c)^{2-\gamma}\,,
\end{equation}
with $\gamma = 1/2$ here. The exponent $\gamma$ is the entropic exponent, and it determines the singular behavior of the free energy (here, the second derivative of the free energy). This singular behavior of the free energy is the hallmark of a phase transition. The nature of this transition can be understood from perturbation theory. The leading Feynman diagrams at large $N$, called cactus diagrams, have a treelike structure typical of branched polymers, and $F_N(g) \approx W(\rho_0)$ yields the exact resummation of these diagrams. The singularity is thus interpreted as the point where the proliferation of these diagrams becomes uncontrollable. In discrete quantum gravity approaches, this point usually corresponds to the continuum limit \cite{di19952d}. Furthermore, the value $\gamma = 1/2$ is typical of branched polymers, which are also found in the leading (melonic) sector of colored random tensors \cite{Bonzom_2011}.

\paragraph{Zinn-Justin-Brézin Renormalization group approach.} The RG approach proposed by Brézin and Zinn-Justin for random matrices, and discussed in the introduction, can also be considered for vector models, as presented by Zinn-Justin in his article \cite{Zinn_Justin_2014}. We briefly present this approach here for a quartic model. Following Zinn-Justin, it is assumed that a variation $N + \delta N$ can be compensated by a change in the couplings, without affecting the continuum limit $N \to \infty$.

Let us therefore consider a quartic model, whose partition function reads:
\begin{equation}
Z_N(g):=\left(\frac{N}{2\pi}\right)^{N/2}\int \, \dd \phi\, e^{-N \left(\frac{1}{2} \phi^2+\frac{g}{4}(\phi^2)^2\right)}\,,\label{modelquartic}
\end{equation}
and we define the potential $V_N(\phi):=\frac{1}{2} \phi^2+\frac{g}{4}(\phi^2)^2$. Following the standard Wilsonian strategy, we will track the evolution of the couplings, starting from the partition function $Z_{N+1}$, by isolating the $N$ smallest components in the integral and integrating over the $(N+1)$-th component. Concretely, we write $\phi=\varphi+\psi$, where $\varphi\in \mathbb{R}^N$ and $\psi\in \mathbb{R}$, which are then identified with the "high" and "low" modes in ordinary Wilsonian RG, so that the partition function reads:
\begin{equation}
Z_{N+1}:=\left(\frac{N+1}{2\pi}\right)^{(N+1)/2}\int \, \dd \varphi\,\int \dd \psi\, e^{-(N+1) \left(\frac{1}{2} \varphi^2+\frac{1}{2}\psi^2+\frac{g}{4}((\varphi^2)^2+2\varphi^2 \psi^2+\psi^4)\right)}\,.
\end{equation}
Assuming $N$ is sufficiently large, the previous integral can also be written as:
\begin{equation}
Z_{N+1} \approx \sqrt{e} \int \dd \varphi e^{-(N+1) V_N(\varphi)}\, \underbrace{\int \frac{\dd \psi}{\sqrt{2\pi N^{-1}}}\, e^{-N \left(\frac{1}{2}\psi^2+\frac{g}{4}\left(2 \varphi^2 \psi^2+\psi^4\right)\right)}}_{:=I_N}\,.
\end{equation}
We will calculate the second integral perturbatively, with the Gaussian measure over $\psi$ being normalized to unity. By expanding the integral $I_N$ in powers of $g$:
\begin{equation}
I_N=\int \frac{\dd \psi}{\sqrt{2\pi N^{-1}}}\, e^{- \frac{N}{2}\psi^2}\left[1-\frac{N g}{4}\left(2 \varphi^2 \psi^2+\psi^4\right)+\frac{N^2 g^2}{32} \left(4 (\varphi^2)^2\psi^4+4 \varphi^2 \psi^6+\psi^8\right)+\cdots\right]\,.
\end{equation}
Using Wick's theorem, we find, besides constants that change the overall normalization of the partition function, the following corrections to the $\varphi$ field interactions at the leading order in $N$:
\begin{equation}
I_N\sim 1 -\frac{g}{2} \varphi^2+\frac{3 g^2}{8} (\varphi^2)^2+\mathcal{O}(1/N)\,,
\end{equation}
or, since $\ln I_N \approx -\frac{g}{2} \varphi^2 + \frac{g^2}{4} (\varphi^2)^2$, we find that the effective potential $N V_{\text{eff},N}$ "at scale $N$" is written as:
\begin{equation}
N V_{\text{eff},N}=(N+1) V_{N+1}- \ln I_N\approx (N+1) V_{N+1}+\frac{g}{2} \varphi^2- \frac{g^2}{4} (\varphi^2)^2\,.
\end{equation}
Interpreting $V_{N+1}$ as the effective potential at scale $N+1$, and using the approximation valid at large $N$:
\begin{equation}
N V_{N+1}-N V_{\text{eff},N} \approx N \frac{\dd}{\dd N} \left( V_{\text{eff},N}\right)\,,
\end{equation}
and therefore, to a first approximation:
\begin{equation}
-N \frac{\dd}{\dd N} V_{\text{eff},N}\approx V_{\text{eff},N}+\frac{g}{2} \varphi^2- \frac{g^2}{4} (\varphi^2)^2\,.\label{floteq1er}
\end{equation}
This equation tells us that the propagator of the $\psi$ field is modified, and the effective potential will therefore have the following expression at the quartic order:
\begin{equation}
V_{\text{eff},N}=\frac{1}{2}m^2(N) \varphi^2+\frac{g(N)}{4} (\varphi^2)^2\,,
\end{equation}
where we have highlighted the explicit $N$-dependence of the effective couplings. The propagator of the $\psi$ mode is therefore $m^{-2}/N$ at an arbitrary step of the flow, so that equation \eqref{floteq1er} is generally written as:
\begin{equation}
\boxed{-N \frac{\dd}{\dd N} V_{\text{eff},N}=V_{\text{eff},N}+\frac{g(N)}{2 m^2(N)} \varphi^2- \frac{g^2(N)}{4 m^4(N)} (\varphi^2)^2\,.}
\end{equation}

The flow equations for the couplings $m^2(N)$ and $g(N)$ are deduced from this immediately:
\begin{align}
-N \frac{\dd m^2}{\dd N}&= m^2+\frac{g}{m^2}\\
-N \frac{\dd g}{\dd N}&= g-\frac{g^2}{m^4}\,.
\end{align}
To compare models sharing the same "infrared" (IR) limit, it is customary to renormalize the field so that the kinetic term remains constant along the flow \cite{Zinn_Justin_2014,eichhorn2013continuum}. This corresponds to the rescaling $\varphi \to m^{-1} \varphi$, and the renormalized coupling constant $\bar{g}$ is defined as:
\begin{equation}
\bar{g}:= m^{-4} g\,,
\end{equation}
which gives us the $\beta$ function:
\begin{equation}
\boxed{\beta(\bar{g})\equiv N\frac{\dd \bar{g}}{\dd N}=\bar{g}+3\bar{g}^2\,.}\label{betavector}
\end{equation}
This equation admits a relevant interacting fixed point at $\bar{g}_* = -1/3$, with a critical exponent $\theta := -\beta^\prime(\bar{g}_*) = 1$. The flow of the relevant effective coupling $\bar{v} \approx \bar{g} - \bar{g}_*$ therefore locally follows the scaling law $v = \bar{v} N$. We saw earlier that the critical value of the quartic model is $g_c = -1/4$, whereas the critical exponent of the sole relevant coupling $a_1$ (see \eqref{Z_Ndoublescal}) was $2/3$. The values obtained  $-1/3$ for the coupling and $1$ for the exponent are therefore in qualitative agreement with the analytical predictions.

This perturbative method can be generalized non-perturbatively within the Wetterich formalism by adding a regulator $\Delta V_{N}$ to the initial potential, for example:
\begin{equation}
\Delta V_{N}=\frac{m^2(N)}{2}\, \sum_{i} \left(\frac{N}{i}-1\right)\theta \left(1-\frac{i}{N}\right) \varphi_i^2\,,\label{regul}
\end{equation}
where $\theta(x)$ is the standard Heaviside function. The regulator function:
\begin{equation}
R_N(i):=m^2(N) \left(\frac{N}{i}-1\right)\theta \left(1-\frac{i}{N}\right)\,,
\end{equation}
is constructed so as to interpolate between two regimes:
\begin{itemize}
    \item An ultraviolet (UV) regime, $N \to \Lambda$ (for some large cut-off $\Lambda$ on the size of the vector), for which $R_{N\to \Lambda} \sim \Lambda$ and quantum fluctuations are frozen.
    \item An IR regime, for which $R_N$ formally vanishes when $N \ll 1$ and all fluctuations are integrated out.
    \item The IR fluctuations relative to the scale $N$, i.e., such that $i \leq N$, acquire a relatively large effective mass (of the order of $N$) and therefore decouple from the flow. Conversely, for the UV modes ($i > N$), the regulator vanishes, and they are simply integrated into the effective action.
\end{itemize} 
The regulator \eqref{regul} incorporates these limits, as well as the Zinn-Justin approach of partial integration over the components of the vector having large label with respect to some cut-off. Applications to random matrices and tensors have been considered in \cite{eichhorn2019status,eichhorn2014towards,eichhorn2018flowing,eichhorn2013continuum}, which the reader can consult.

To conclude, let us summarize some of the issues posed by this method, an extensive discussion of which can be found in \cite{gurau2024quantum}:

\begin{enumerate}
    \item Explicit symmetry breaking: The strict $O(N)$ invariance of the vector model does not single out any "fast" or "slow" modes, which is manifest from the fact that the free propagator is proportional to the identity matrix. By choosing a partial integration order to construct the Wilsonian flow, one effectively breaks $O(N)$ invariance. This can lead to consequences that are difficult to control, particularly for random matrices and tensors\footnote{The case of vectors is special, in the sense that Ward identities only yield contributions at order $1/N$, see \cite{Natta:2024nke}.} (modifications of Ward identities, proliferation of "derivative" interactions whose couplings depend on matrix indices, such as $\sum_{i,j,k,l} i^q j^m M_{ij}M_{jk}M_{kl} M_{lm}$, etc.), so that the restoration of symmetry in the formal "IR" limit $N \to 0$ is not guaranteed \cite{Lahoche:2020pjo,Lahoche:2019ocf}.
    \item The power-counting issue: Power counting determines a semblance of canonical dimension (corresponding, for example, to the linear term in \eqref{betavector}). However, all couplings are irrelevant from this perspective, which makes the discussion of vertex expansion convergence more subtle, a complication further exacerbated by the presence of "derivative" interactions.
    \item The inconsistency of the infrared limit: In this formalism, the IR limit corresponds to the formal limit $N \to 0$. However, this limit directly conflicts with the "large $N$" approximation used to derive the flow equations, since the existence of fixed points is only guaranteed in that limit.
\end{enumerate}
In this paper, we present a method that resolves all of these points, notably by decoupling the problem of the IR limit from that of the continuum limit, $N \to \infty$.

\section{Relational background and the continuum limit}\label{relationalback}

In this section, we present the relational background field formalism for random vectors. We will take the example of a quartic model to simplify the presentation, but the method is not limited to these models alone, see also the discussion in Section \ref{conclusion}. The construction provides a canonical notion of scale and establishes a correspondence between the vector model and a three-dimensional effective non-local field theory in the deep IR, as explained at the end of this section.

\subsection{Partial intermediate field vacuum}

We consider a quartic vector model whose partition function is given by \eqref{modelquartic}. The potential reads:
\begin{equation}
V(\phi^2)=\frac{1}{2}\, \phi^2+\frac{g}{4}\, (\phi^2)^2\,, \label{quartic1}
\end{equation}
The Hubbard–Stratonovich transformation \cite{Hubbard:1959ub,lionni2016intermediatefieldrepresentationpositive} is a method that replaces the quartic interaction with a three-point interaction, coupling two $\phi$ fields to an auxiliary field associated with a Gaussian measure. Integrating over the $\phi$ field, which has been rendered Gaussian, leads to an effective theory for the intermediate field, which proves valuable in certain approaches of constructive field theory \cite{rivasseau2009constructive}. In the case of the vector field, this intermediate field can be of a scalar nature — leading to a three-point interaction of the form $\sigma \phi^2$ — or of a matrix nature — leading to a three-point interaction of the form $\sum_{i,j} \sigma_{ij} \phi_i \phi_j$, where $\sigma_{ij}$ is a \textit{real and symmetric} matrix. The first choice involves only a single degree of freedom and is not particularly interesting from the point of view of scale structuring. The second choice involves a large number of degrees of freedom, which, as we shall see, can condense into a form largely decoupled from the remaining degrees of freedom.

A relational background field approach consists in decomposing only a part of the quartic interaction, by replacing the potential \eqref{quartic1} with the equivalent two-field potential:
\begin{equation}
V(\phi^2,\sigma)=\frac{1}{2}\,\phi^2+\frac{1}{4}\, \Tr \sigma^2+\frac{\beta}{2}\, \sum_{i,j} \sigma_{ij} \phi_i \phi_j+\frac{g^\prime}{4}\, (\phi^2)^2\,.\label{quartic2} 
\end{equation}
The equivalence with model \eqref{quartic1} can be established by integrating over the $\sigma$ field (with the flat Lebesgue measure $\dd\sigma:=\prod_{i\leq j}\dd\sigma_{ij}$, which is invariant under orthogonal conjugation), or, equivalently in this case, by evaluating the potential at the saddle point:
\begin{equation}
\frac{\partial}{\partial \sigma_{ij}}\, V(\phi^2,\sigma)=0\,,
\end{equation}
which leads to the condition:
\begin{equation}
\frac{1}{2} \sigma_{ij}+\frac{\beta}{2} \phi_i\phi_j=0\,.
\end{equation}
By substituting this condition back into the potential \eqref{quartic2}, we obtain:
\begin{align}
V(\phi^2,\sigma)\bigg\vert_{\text{saddle}}&=\frac{1}{2}\,\phi^2+\frac{\beta^2}{4}\, (\phi^2)^2-\frac{\beta^2}{2}\, (\phi^2)^2+\frac{g^\prime}{4}\, (\phi^2)^2\\
&=\frac{1}{2}\,\phi^2+\frac{g^\prime-\beta^2}{4}\, (\phi^2)^2\,,
\end{align}
so that the two models are equivalent provided that:
\begin{equation}
\boxed{g^\prime-\beta^2\equiv g\,.}\label{defgprime}
\end{equation}
The parameter $\beta$ controls the "splitting" of the $\sigma$ background within the set of initial degrees of freedom of the $\phi$ field. We will see that this splitting leads to a non-trivial result in the case where this background field is matrix-valued.

Let us begin by analyzing the problem from a probabilistic perspective. Let $P[\phi]$ be the probability of a $\phi$ configuration, given by \eqref{probaphi}, and let $Q(A)$ be the probability density of the real symmetric matrix $A$ with entries $A_{ij}$. The latter is assumed to be $O(N)$-invariant (Gaussian orthogonal ensemble (GOE)) and given by:
\begin{equation}
Q(A):=\frac{1}{Z_A}\,\exp \left(-\frac{N}{4}\, \Tr A^2\right)\,,\quad Z_A:=\int \dd A\, \exp \left(-\frac{N}{4}\, \Tr A^2\right)\,,
\end{equation}
where $\dd A$ is the (flat) Lebesgue measure on symmetric matrices, such that $\dd (O^T A O) = \dd A$ for any orthogonal matrix $O$, with $T$ denoting the ordinary transpose here. The joint probability for the variable $(A, \phi)$ is therefore written as $P(\phi) Q(A)$. Let us now consider the matrix
\begin{equation}
\boxed{\sigma:=A-\beta \phi \phi^T\,,}\label{defsigma}
\end{equation}
with $\beta \in \mathbb{R}$. The probability law $R(\sigma)$ for this new variable becomes:
\begin{align}
\nonumber R(\sigma)&:=\int \dd A \int \dd \phi\, P(\phi) Q(A) \delta (\sigma-A+\beta \phi \phi^T)\\\nonumber
&= \int \dd \phi \,\left(\frac{N}{2\pi}\right)^{N/2}\, \frac{e^{-N V(\phi^2)}e^{-\frac{N}{4}\Tr (\sigma+\beta \phi \phi^T)^2}}{Z_A Z_N(g)}\\
&=\int \dd \phi \,\left(\frac{N}{2\pi}\right)^{N/2}\, \frac{e^{-N \left(\frac{1}{2}\,\phi^2+\frac{1}{4}\, \Tr \sigma^2+\frac{\beta}{2}\, \sum_{i,j} \sigma_{ij} \phi_i \phi_j+\frac{g^\prime}{4}\, (\phi^2)^2\right)}}{Z_A Z_N(g)}\,,
\end{align}
The associated partition function,
\begin{equation}
Z_R:=Z_A Z_N(g) \equiv \int \dd \sigma \int \dd \phi \,\left(\frac{N}{2\pi}\right)^{N/2}\, e^{-N \left(\frac{1}{2}\,\phi^2+\frac{1}{4}\, \Tr \sigma^2+\frac{\beta}{2}\, \sum_{i,j} \sigma_{ij} \phi_i \phi_j+\frac{g^\prime}{4}\, (\phi^2)^2\right)}\,,
\end{equation}
can also be identified, up to a constant factor, with the initial model, provided that $g^\prime$ satisfies \eqref{defgprime}. We thus obtain the following result:
\begin{claim}
Non-perturbatively, the intermediate field $\sigma$ is the free sum of a Gaussian random matrix whose off-diagonal elements have variance $1/N$, translated by the rank $1$ random perturbation $-\beta \phi {\phi}^T$.
\end{claim}
Let us now analyze the behavior of the theory in the limit $N \to \infty$. For the vector sector, concentration of the measure implies that the fluctuations of $\phi^2 \equiv \rho$ become negligible, as discussed in the previous section. For the matrix sector, random matrix theory similarly shows that the measure of the matrix $A$ concentrates, while its spectrum converges to a deterministic distribution given by Wigner's theorem, which can be written as \cite{potters2020first}:
\begin{theorem}\label{th0}
Let $M$ be a symmetric random matrix with independent, centered Gaussian entries with off-diagonal variance $\nu^2/N$. Let $\{\lambda_i\}$ be its $N$ eigenvalues, $i=1,2,\cdots, N$. Then, as $N\to \infty$, the empirical eigenvalue distribution $\mu_{\text{E}}(\lambda):=\frac{1}{N}\sum_{i=1}^N\delta(\lambda-\lambda_i)$ converges weakly to the semicircle law $\mu(\lambda)$:

\begin{equation}
\mu_{\text{E}}(\lambda) \to \mu(\lambda) := \left\{
    \begin{array}{ll}
        \frac{\sqrt{4\nu^2-\lambda^2}}{2\pi \nu^2} & \mbox{if}\,  -2\nu \leq \lambda \leq 2\nu \\
        0 & \mbox{if} \quad  \lambda^2 > 4\nu^2\,.
    \end{array}
\right.
\end{equation}
\end{theorem}
A matrix of this type will be called a \textit{Wigner matrix}, and Figure \ref{figWig} illustrates the theorem numerically.

Formally, the sum \eqref{defsigma} is a rank-1 perturbation of a Wigner matrix, which is typically encountered in signal processing theory\footnote{In this context, matrix $A$ represents the noise, $-\beta \phi \phi^T$ the information-bearing signal, and $\sigma$ the empirical data.} \cite{bloemendal2013limits}, and we have the following theorem, originally established by Baik, Ben Arous, and Péché (BBP) for sample covariance matrices \cite{baik2005phase} and extended to finite-rank deformations of Wigner matrices in \cite{peche2006largest,feral2007largest,capitaine2009largest}:
\begin{figure}
\begin{center}
\includegraphics[scale=0.5]{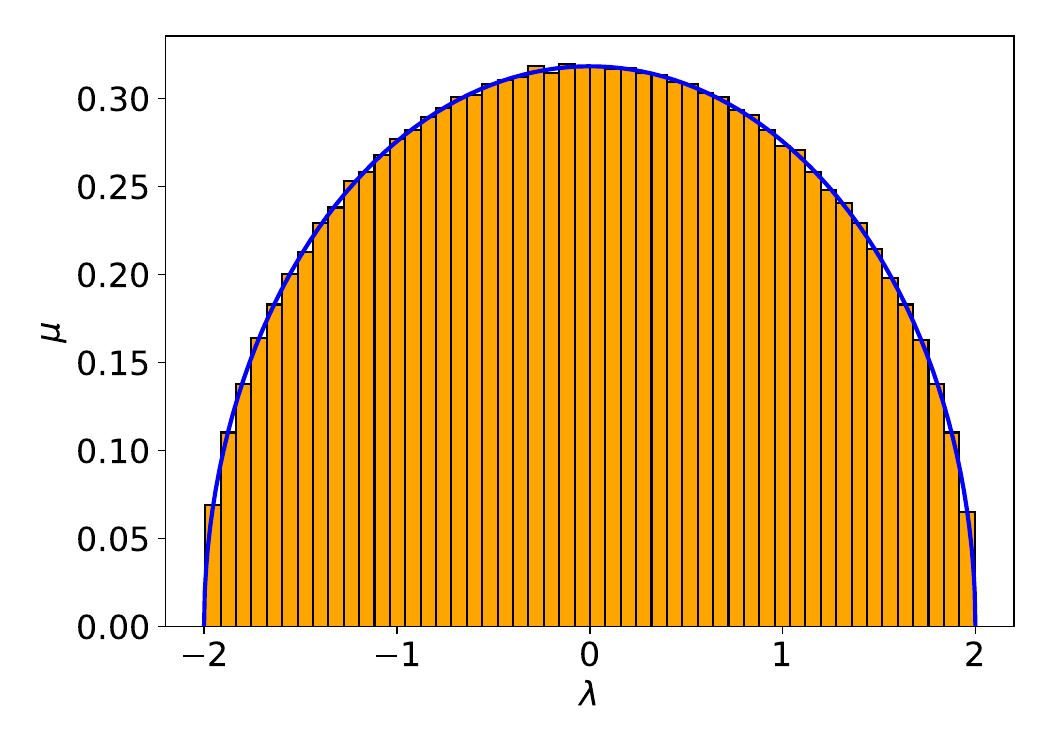}
\end{center}
\caption{ Illustration of the convergence toward Wigner's law.
The histogram shows the eigenvalue distribution for a Wigner matrix of size $10^4$ with variance $1/N$.
The blue line represents the limiting analytical Wigner semicircle law.}\label{figWig}
\end{figure}

\begin{theorem}\label{thBBP}
  Let $M$ be a Wigner matrix of variance $1/N$ drawn from the Gaussian Orthogonal Ensemble (GOE), and let $Q=\alpha uu^{\mathsf{T}}+M$ be a spiked Wigner matrix with $u^{\mathsf{T}}u=1$.
We have:
\begin{itemize}
    \item For $\vert \alpha \vert \leq 1$, the largest and smallest eigenvalues of $Q$ converge almost surely to $2$ and $-2$ respectively as $N\to \infty$. Their fluctuations are of order $N^{-2/3}$ with Tracy--Widom limiting distributions.
    \item For $\vert \alpha \vert > 1$, the extreme eigenvalue associated with the sign of $\alpha$ (the largest if $\alpha > 1$, the smallest if $\alpha < -1$) separates from the bulk. It converges almost surely to $\alpha+1/\alpha$ as $N\to \infty$, and its fluctuations are of order $N^{-1/2}$ with a Gaussian limiting distribution. The other extreme eigenvalue remains at the bulk edge ($2$ or $-2$).
\end{itemize}
\end{theorem}
An elementary proof can be found in \cite{potters2020first} or in Appendix F of \cite{finotello2026datafieldtheorytheory}. With the definition of $\rho \equiv \phi^2$ given in the previous section, one can write $\phi = \sqrt{\rho} u$, with $u^2 = 1$, so that the definition of $\sigma$ is also $\sigma = A - \beta \rho u u^T$. The BBP theorem therefore implies the following result:
\begin{claim}
In the limit $N \to \infty$, $\sigma$ is a Wigner matrix with off-diagonal variance $1/N$ if:
\begin{equation}
\beta \rho_0 \leq 1\,.
\end{equation}
If this condition is satisfied, the spectrum of $\sigma$ is a Wigner semicircle.
\end{claim}

Thus, if the preceding condition is satisfied by $\rho_0$ at the saddle point, then at large $N$, the initial model is equivalent to the $p=2$ soft spin model \cite{de2006random}:
\begin{equation}
Z_R \sim\left(\frac{N}{2\pi}\right)^{N/2}\, \left[ \int \dd \phi \, e^{-N \left(\frac{1}{2}\,\phi^2+\frac{\beta}{2}\, \sum_{i,j} \sigma_{ij} \phi_i \phi_j+\frac{g^\prime}{4}\, (\phi^2)^2\right)}\right]_{\sigma\in\mathrm{GOE}}\,,\label{modeleff}
\end{equation}
where the matrix $\sigma$ is interpreted as disorder\footnote{In other words, when $\beta \rho_0 \leq 1$, "quenched=annealed".}. The rotational invariance of the measure of $\sigma$, and that of the $\phi$ field, allows fixing the gauge in the basis where $\sigma$ is diagonal, which provides a natural basis to organize the modes. In this basis, the effective, free propagator of the $\phi$ field:
\begin{equation}
C(\lambda):=\frac{1}{N}\frac{1}{1+\beta \lambda}\,.
\end{equation}
The "freezing" of the background field $\sigma$ onto a Wigner matrix, together with the gauge fixing, endows the residual fluctuations of the $\phi$ field with a scale through the now non-trivial spectrum of the propagator. This scale makes it possible to define a Wilsonian RG by introducing a natural ordering for the partial integration of the field modes in the basis in which $\sigma$ is diagonal. The initial gauge invariance is never explicitly broken and is instead encoded through Ward identities. In the vector model considered here, these identities do not contribute at leading order in the continuum limit $N \to \infty$, as will be discussed in section \ref{conclusion}.

\subsection{Equivalence with the original model}

Although the BBP theorem guarantees the equivalence described above, it is instructive to verify this equivalence directly at the saddle point. We introduce the identity into the integral:
\begin{equation}
1=\int \dd \rho \, \delta(\rho-\phi^2)\,,
\end{equation}
we obtain:
\begin{equation}
Z_R \sim \left[ \int \dd \phi \int \dd \rho \, e^{-N \left(\frac{1}{2}\,\rho+\frac{\beta}{2}\, \sum_{i,j} \sigma_{ij} \phi_i \phi_j+\frac{g^\prime}{4}\, \rho^2\right)} \delta(\rho-\phi^2)\right]_{\sigma\in\mathrm{GOE}}\,,\label{modeleff2}
\end{equation}
It will be convenient to work in the basis where $\sigma$ is diagonal. Let $u_i^{(\lambda)}$ denote an eigenvector of $\sigma$ with eigenvalue $\lambda$, such that $\sum_j \sigma_{ij} u_j^{(\lambda)}=\lambda\, u_i^{(\lambda)}$. By defining:
\begin{equation}
\phi_\lambda:=\sum_{i=1}^N\, \phi_i u_i^{(\lambda)}\,,
\end{equation}
Using the following integral representation of the Dirac $\delta$ function\footnote{The integral contour is parallel to the imaginary axis, with the real part of $z$ fixed at $z_0$. The factor $N$ simply introduces an overall rescaling.}:
\begin{equation}
\delta(\rho-\phi^2)=\frac{N}{4i\pi}\int_{z_0-i \infty}^{z_0+i\infty} \dd z\, e^{\frac{Nz}{2} (\rho-\phi^2)}\,,
\end{equation}
we obtain, by computing the Gaussian integral over $\phi$:
\begin{equation}
Z_R \sim \left[  \int \dd \rho \int_{z_0-i \infty}^{z_0+i\infty} \dd z \, e^{-N \left(\frac{1}{2}(1-z)\,\rho+\frac{1}{2N}\, \sum_{\lambda} \log (z+\beta \lambda)+\frac{g^\prime}{4}\, \rho^2\right)}\right]_{\sigma\in\mathrm{GOE}}\,.\label{modeleff3}
\end{equation}
The saddle-point equations are obtained by setting the derivatives with respect to $\rho$ and $z$ to zero:
\begin{align}
\frac{1}{2}(1-z)+\frac{1}{2}g^\prime \rho&=0\\
-\frac{1}{2} \rho +\frac{1}{2N} \sum_{\lambda=1}^N\, \frac{1}{z+\beta \lambda}&=0\,.
\end{align}
The first saddle-point equation gives $z=1+g^\prime \rho$, and substituting this back into the second yields:
\begin{equation}
\rho_0=\frac{1}{N}\sum_{\lambda=1}^N\, \frac{1}{1+g^\prime \rho_0+\beta \lambda}\,.
\end{equation}
In the limit $N \to \infty$, Wigner’s theorem allows us to replace the normalized sum over eigenvalues by an integral against the semicircle distribution: $\sum_\lambda f(\lambda)\simeq N \int_{-2}^{+2}\, \mu(\lambda) f(\lambda)$, where $\mu(\lambda)$ is the semicircle distribution. The saddle-point equation therefore becomes, in the limit $N \to \infty$:
\begin{equation}
\boxed{\rho=\int_{-2}^{+2}\, \dd \lambda\, \frac{\mu(\lambda)}{1+g^\prime \rho+\beta \lambda}\,.}
\end{equation}
The integral can be calculated analytically\footnote{The integral can also be obtained from the Stieltjes transform of the matrix $\beta \sigma$.}, provided that the denominator does not vanish on the support of the semicircle distribution, i.e. $1+g^\prime \rho_0 \geq 2\beta$. We find:
\begin{equation}
\rho_0=\frac{(g^\prime \rho_0 +1) \left(1-\sqrt{1-\frac{4 \beta ^2}{(g^\prime \rho_0 +1)^2}}\right)}{2 \beta ^2}\,.
\end{equation}
This equation can be rewritten:
\begin{equation}
(2\beta^2 \rho_0 - g^\prime \rho_0 -1)^2 =  \left((g^\prime \rho_0+1)^2-4 \beta ^2\right)\,,
\end{equation}
that is:
\begin{equation}
(g^\prime-\beta^2)\rho^2_0 + \rho_0 -1 =  0\,,
\end{equation}
and we recognize the saddle-point equation \eqref{condition_sadle_quart} of the $O(N)$ model, provided that $g=g^\prime-\beta^2$, as expected. Furthermore, by expressing $1+g^\prime \rho_0=\rho_0^{-1}(1+\beta^2 \rho_0^2)$ from the previous equation, the condition $1+g^\prime \rho_0 \geq 2\beta$ can also be written as $(\beta \rho_0-1)^2\geq 0$. The bound is therefore reached for $\beta \rho_0=1$, as expected from the BBP theorem.

\begin{figure}
\begin{center}
\includegraphics[scale=0.6]{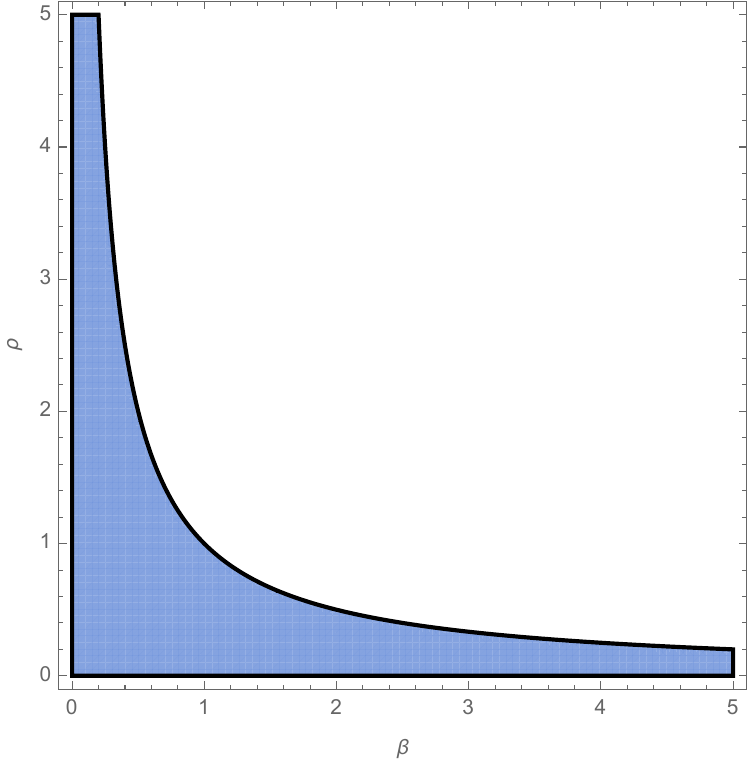}
\end{center}
\caption{The validity domain of the equivalence between quenched and annealed limit.}\label{figdomaine}
\end{figure}

{
\begin{remark}
It is easy to check that the saddle-point free energy equals that of the $O(N)$ model, up to some irrelevant constant. Indeed, the saddle point for the original model gives:
\begin{equation}
W(\rho_0) = \frac{1}{4}\rho_0 + \frac{1}{4} - \frac{1}{2}\ln \rho_0\,,\label{saddleP}
\end{equation}
and for the effective "R" model:
\begin{equation}
W_{\text{R}}(\rho_0, z_0) = -\frac{g'}{4}\rho_0^2 + \frac{1}{2} \int \mu(\lambda) \ln(1+g'\rho_0+\beta\lambda) d\lambda\,.
\end{equation}
To compute the integral, we can integrate the Stieltjes transform with respect to the variable $A = 1+g'\rho_0$, leading to:
\begin{equation}
\int \mu(\lambda) \ln(A+\beta\lambda) d\lambda = \frac{A^2 - A\sqrt{A^2-4\beta^2}}{4\beta^2} + \ln(A+\sqrt{A^2-4\beta^2}) + C\,.
\end{equation}
Using $A - \sqrt{A^2-4\beta^2} = 2\beta^2 \rho_0$, and after some algebraic manipulations, we recover \eqref{saddleP}.
\end{remark}
}
\begin{remark}

The condition $(\beta \rho_0-1)^2\geq 0$ is satisfied for any value of $\beta$ and $\rho_0$; the BBP bound is nevertheless already encoded in the choice of the branch of the square root. Indeed, the explicit solution requires $\sqrt{A^2-4\beta^2}=A-2\beta^2\rho_0\geq 0$, with $A:=1+g^\prime\rho_0=\rho_0^{-1}(1+\beta^2\rho_0^2)$, i.e. $1-\beta^2\rho_0^2\geq 0$, which is precisely $\beta\rho_0\leq 1$. When $\beta \rho_0 > 1$, the saddle-point approximation breaks down, with a significant (macroscopic) part of the degrees of freedom condensing onto the isolated eigenvalue emerging from the bulk, see \cite{de2006random,Lahoche:2024huc}. In this regime, the correspondence with the original $O(N)$ model is lost. Figure \ref{figdomaine} represents the precise domain where this correspondence is valid, bounded by the hyperbola $\rho=1/\beta$.
\end{remark}

To conclude, let us verify that the condition $\beta \rho_0 \leq 1$ includes the critical condition $W^{\prime\prime}(\rho)=0$. For the quartic model, this condition is written as $g \rho_0^2=-1$ (see equation \eqref{critical_cond}). Substituting this into the saddle-point condition \eqref{condition_sadle_quart} gives $\rho_0=2$, and we must therefore have $2 \beta \leq 1$, leading to the following result:
\begin{claim}
Up to and including the critical domain, there exists a finite domain:
\begin{equation}
\boxed{0\leq \beta \leq \frac{1}{2}\,.}
\end{equation}
such that the spectrum of $\sigma$ is frozen, in the limit $N \to \infty$, onto the semicircle law.
\end{claim}
In other words, the quenched model \eqref{modeleff} contains the critical point of the original $O(N)$ model. Note, however, that at the critical point $g_c = -1/4$, the relation \eqref{defgprime} implies $g' = -1/4 + \beta^2 < 0$ for all $\beta \in (0, 1/2)$, while at the boundary $\beta = 1/2$ both $g'$ and the bare mass $m = \beta^{-1} - 2$ introduced below vanish simultaneously, reducing the residual theory to a free Gaussian model.

\subsection{Non-local effective field theory}

Before implementing a coarse-graining procedure based on the eigenvalues of the effective propagator, we can give a more suggestive form to the limiting effective theory \eqref{modeleff}. Let us recall the following definition here:
\begin{definition}
Let $K$ be a bounded linear operator between two normed spaces $K: U\to V$. The \textit{operator norm} is defined by $\Vert K \Vert_{\text{op}}:=\sup_{v\in U, \Vert v \Vert_U=1} \,\Vert K v \Vert_V$, and for an $N\times N$ real and symmetric matrix, $\Vert K \Vert_{\text{op}}=\max_{1\leq i \leq N} \vert \lambda_i \vert$, i.e., the spectral radius.
\end{definition}
E.g. for a Wigner matrix whose off-diagonal elements have variance $\nu^2/N$, the operator norm tends to $2\nu$ in the limit $N \to \infty$. 

We will therefore define the \textit{generalized momentum} $p$ such that, at finite $N$:
\begin{equation}
p:=\lambda+\Vert \sigma \Vert_{\text{op}}\,.
\end{equation}
By construction, $p$ is non-negative, and, in the limit $N \to \infty$:
\begin{equation}
p\equiv \lambda+2\nu\,.
\end{equation}
In the remainder, and as before, $\nu=1$. Furthermore, the Wigner distribution for $\lambda$, $\mu(\lambda)$, induces a distribution $\rho(p)$ for the generalized momentum:
\begin{equation}
\rho(p):=\frac{\sqrt{p(4-p)}}{2\pi}\,.
\end{equation}
It will also be convenient to rescale the field $\phi \to \beta^{-1/2} \phi$, and to introduce the \textit{bare mass} $m$, defined in the continuum limit by:
\begin{equation}
m:=\beta^{-1}-2\,,
\end{equation}
so that the new propagator is written as:
\begin{equation}
C(p):=\frac{1}{N}\frac{1}{p+m}\,,
\end{equation}
and furthermore, it is easy to verify that, for $\beta < 1/2$, $m > 0$ and the propagator is never singular over the entire domain\footnote{Or, more rigorously stated, the probability that $1+\beta \lambda$ is negative goes to zero as $N\to \infty$, as $\beta < 1/2$.} $p \in [0,4]$. Expressed in these variables, and defining $\Phi(p):=N\phi_{\lambda=p-2}$, the partition function takes the following suggestive form, valid in the continuum limit:
\begin{equation}
Z_N \sim \int [\dd \Phi(p)]\, e^{-\mathcal{S}[\Phi]}\,,
\end{equation}
with $[\dd \Phi(p)]:=\prod_p \dd \Phi(p)$ and, for the quartic model we are considering:
\begin{equation}
\mathcal{S}[\Phi]:=\frac{1}{2} \int \rho(p) \Phi(p) (p+m)  \Phi(p)+ \frac{u}{4N} (\Phi \cdot \Phi)^2 \,,\label{classicalaction}
\end{equation}
where we have defined $u:=g^\prime/\beta^2$, the factor $\beta^{-2}$ coming from the field rescaling $\phi\to\beta^{-1/2}\phi$ (see also \eqref{exact_bare_params}), and with the dot product defined by:
\begin{equation}
\Phi \cdot \Phi:=\int_0^{4}\, \rho(p)\, \Phi^2(p)\,.
\end{equation}
In this form, in the continuum limit, the theory takes the form of a non-local field theory, characterized by an unconventional distribution of the generalized momentum. In this limit, the field fluctuations acquire, around the Gaussian fixed point, a non-trivial scale (given by the propagator), which defines canonical notions of ultraviolet (UV) and IR:
\begin{definition}
The infrared regime corresponds to generalized momenta near the lower spectral edge, $p \ll 1$, and conversely, the ultraviolet regime corresponds to momenta near the upper end of the spectral support, $p = \mathcal{O}(1)$. It is worth noting that the theory is also characterized by the existence of a fundamental UV cutoff $\Lambda =4$.
\end{definition}
The IR limit is particularly interesting, given that the distribution $\rho(p)$ follows a power law:
\begin{equation}
\rho(p) \sim \sqrt{p}\,.
\end{equation}
In an ordinary Euclidean field theory on $\mathbb{R}^D$, the distribution of Fourier momenta follows the law $(\vec{p}\,^2)^{\frac{D-2}{2}}$; the asymptotic behavior of the effective theory therefore resembles that of a non-local field theory in dimension $3$.

%Discussion du champ de fond ordinaire
\section{Renormalization group and effective scaling}\label{renormalization}

In this section, we present an elementary, perturbative approach to the Wilsonian RG constructible from the notion of scale that we have brought out.

\subsection{Wilsonian flow}

Wilsonian RG \cite{Wilson:1972zzb,Zinn-Justin:2019jix,zinn2021quantum} is characterized above all by a specific procedure of partial integration of UV modes, generating, step by step, a series of effective theories sharing the same IR limit. An overview of the general strategy has already been given in Section \ref{summary}; we will therefore only review the essential steps here.

Following the general strategy, we decompose the field $\Phi(p)$ into IR modes according to a certain cutoff $0 \leq \mu \leq 4$ by defining $\Phi_\mu(p)$ such that:
\begin{equation}
\Phi_\mu(p) = \left\{
    \begin{array}{ll}
        \Phi(p) & \mbox{if}\, p\leq \mu \\
        \,0 & \mbox{otherwise.}
    \end{array}
\right.
\end{equation}
We will also define the complementary field $\bar{\Phi}_\mu(p) := \Phi(p) - \Phi_\mu(p)$, such that:
\begin{equation}
\bar{\Phi}_\mu \cdot \Phi_\mu = 0\,.
\end{equation}
The kinetic part of the action \eqref{classicalaction} is therefore written as:
\begin{equation}
\mathcal{S}_{\text{kin}}[\Phi]:=\frac{1}{2} \int \rho(p) \Phi_\mu(p) (p+m)  \Phi_\mu(p)+\frac{1}{2} \int \rho(p) \bar{\Phi}_\mu(p) (p+m)  \bar{\Phi}_\mu(p)\,,
\end{equation}
and the partition function is written, up to constants, as:
\begin{equation}
Z_N \sim \int \dd \nu(\Phi_\mu) \int \dd \nu(\bar{\Phi}_\mu) \, e^{-\frac{u}{4N} \left((\Phi_\mu \cdot \Phi_\mu)^2+(\bar{\Phi}_\mu \cdot \bar{\Phi}_\mu)^2+2 (\Phi_\mu \cdot \Phi_\mu)(\bar{\Phi}_\mu \cdot \bar{\Phi}_\mu)\right)}\,,
\end{equation}
where we have defined the normalized Gaussian measures:
\begin{align}
\dd \nu(\Phi_\mu)&:= \frac{e^{-\frac{1}{2} \int \rho(p) \Phi_\mu(p) (p+m)  \Phi_\mu(p)}}{\int [\dd \Phi_\mu(p)] e^{-\frac{1}{2} \int \rho(p) \Phi_\mu(p) (p+m)  \Phi_\mu(p)}}\,  [\dd \Phi_\mu(p)]\,,\\
\dd \nu(\bar{\Phi}_\mu)&:= \frac{e^{-\frac{1}{2} \int \rho(p) \bar{\Phi}_\mu(p) (p+m)  \bar{\Phi}_\mu(p)}}{\int [\dd \bar{\Phi}_\mu(p)] e^{-\frac{1}{2} \int \rho(p) \bar{\Phi}_\mu(p) (p+m)  \bar{\Phi}_\mu(p)}}\,  [\dd \bar{\Phi}_\mu(p)]\,.
\end{align}
Formally, the integration of the "high" modes $\bar{\Phi}_\mu$ transforms the effective action of the low modes $\Phi_\mu$ by modifying the coupling constants. We will evaluate this transformation using perturbation theory by expanding the exponential in powers of $u$. Since the one-loop contributions stop at order $u^2$, we have:
\begin{equation}
Z_N \sim  \int \dd \nu(\Phi_\mu) e^{-\frac{u}{4N} (\Phi_\mu \cdot \Phi_\mu)^2} \mathcal{I}_N\,,
\end{equation}
with, up to irrelevant contributions:
\begin{align}
\nonumber \mathcal{I}_N&:=\int \dd \nu(\bar{\Phi}_\mu) \, e^{-\frac{u}{4N} \left((\bar{\Phi}_\mu \cdot \bar{\Phi}_\mu)^2+2 (\Phi_\mu \cdot \Phi_\mu)(\bar{\Phi}_\mu \cdot \bar{\Phi}_\mu)\right)}\\
&=\int \dd \nu(\bar{\Phi}_\mu) \, \left(1-\frac{u}{2N}(\Phi_\mu \cdot \Phi_\mu)(\bar{\Phi}_\mu \cdot \bar{\Phi}_\mu)+\frac{u^2}{8 N^2} (\Phi_\mu \cdot \Phi_\mu)^2(\bar{\Phi}_\mu \cdot \bar{\Phi}_\mu)^2\right) +\mathcal{O}(u^3)\,.
\end{align}
The integral is evaluated using Wick's theorem, and we find:
\begin{align}
\int \dd \nu(\bar{\Phi}_\mu) \, (\bar{\Phi}_\mu \cdot \bar{\Phi}_\mu)&= N \int_\mu^4\, \dd p \frac{\rho(p)}{p+m}\\
\int \dd \nu(\bar{\Phi}_\mu) \, (\bar{\Phi}_\mu \cdot \bar{\Phi}_\mu)^2&= N^2 \left(\int_\mu^4\, \dd p \frac{\rho(p)}{p+m}\right)^2+2 N \int_\mu^4\, \dd p \frac{\rho(p)}{(p+m)^2} \,.
\end{align}
Taking the logarithm $\ln \mathcal{I}_N$, the disconnected contributions cancel out at second order, and we find:
\begin{equation}
\ln \mathcal{I}_N = -\frac{u}{2} (\Phi_\mu \cdot \Phi_\mu) \int_\mu^4\, \dd p \frac{\rho(p)}{p+m}+\frac{u^2}{4N}\,(\Phi_\mu \cdot \Phi_\mu)^2\, \int_\mu^4\, \dd p \frac{\rho(p)}{(p+m)^2}+\mathcal{O}(u^3) \,.
\end{equation}
Thus, the couplings $m$ and $u$ for $\Phi_\mu$ are modified as follows:
\begin{align}
m^\prime &=m+u \, \int_\mu^4\dd p \frac{\rho(p)}{p+m}\,,\\
u^\prime &=u-u^2\int_\mu^4\, \dd p \frac{\rho(p)}{(p+m)^2}\,.
\end{align}
Note that these transformations do not entail any change in the field normalization. This is a characteristic feature of vector models, due to the fact that no external face runs inside the loop of the $2$-point function for the leading large-$N$ contribution. Matrix and tensor models, by contrast, do require wave function renormalization \cite{Lahoche:2020pjo,Lahoche_2026_background}.

In general, integrating the modes over the interval $[\mu', \mu]$ yields:
\begin{align}
m(\mu^\prime) &=m(\mu)+u(\mu) \, \int_{\mu^\prime}^\mu\,\dd p \frac{\rho(p)}{p+m(\mu)}\,,\\
u(\mu^\prime) &=u(\mu)-u^2(\mu)\int_{\mu^\prime}^\mu\, \dd p \frac{\rho(p)}{(p+m(\mu))^2}\,,
\end{align}
which can be translated into an infinitesimal version as follows:
\begin{align}
\mu \frac{\dd m}{\dd \mu}&= -u(\mu) \frac{\mu\, \rho(\mu)}{\mu+m(\mu)}\\
\mu \frac{\dd u}{\dd \mu}& = u^2(\mu) \frac{\mu\, \rho(\mu)}{(\mu+m(\mu))^2}
\end{align}

\subsection{Scaling, dimensions and asymptotic fixed points}

The preceding equations describe how the couplings evolve under a scale transformation. Typically, the search for fixed points requires introducing dimensionless couplings. In the present case, however, there is no external notion of scale from which to assign canonical dimensions to the couplings. Nevertheless, an intrinsic notion of dimension emerges within the RG, associated with the explicit scale dependence on $\mu$ appearing on the right-hand side of the equation. Usually, the transition to dimensionless couplings transfers the effect of this scale dependence into the linear term of the flow equations. We generalize this notion of dimension by rescaling the couplings in such a way as to similarly transfer their scale dependence into the linear term. It is worth emphasizing that this approach to dimensionality is not specific to the present context. It has become common in quantum gravity \cite{lahoche2020pedagogical,benedetti2015functional,benedetti2016functional} and has also been employed in several other contexts in recent publications; see \cite{Lahoche:2024gal,finotello2026datafieldtheorytheory,Natta:2024nke} and references therein.

\begin{figure}
\begin{center}
\includegraphics[scale=0.7]{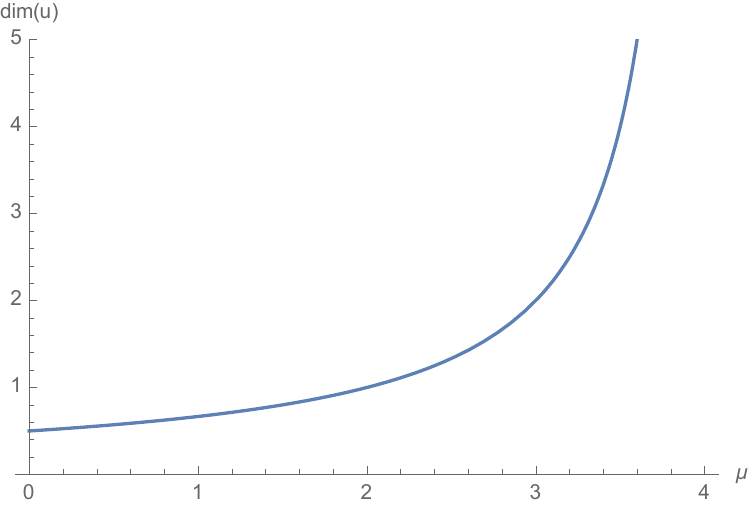}
\end{center}
\caption{Behavior of the canonical dimension for the coupling $u$. It goes toward $1/2$ as $\mu \to 0$.}\label{figdim}
\end{figure}

The transformation:
\begin{equation}
m(\mu)=: \mu \bar{m}(\mu)\,,\qquad u(\mu)=:\mu \rho^{-1}(\mu) \bar{u}(\mu)\,,
\end{equation}
gives us the following equations:
\begin{align}
\boxed{\beta_m=-\bar{m}-\frac{\bar{u}}{1+\bar{m}}\,,\qquad \beta_u=-\dim (u) \bar{u}+\frac{\bar{u}^2}{(1+\bar{m})^2}\,,}
\end{align}
with the definition:
\begin{equation}
\boxed{\dim (u)=- \mu \frac{\dd }{\dd \mu} \ln (\mu^{-1} \rho(\mu))\,.}\label{canonicaldim}
\end{equation}
and $\beta_X:=\dd \bar{X}/\dd \ln \mu$. The function \eqref{canonicaldim} gives the effective dimension, which depends on the scale here. Let us note once again that such a dependence is not new in itself, and has already been encountered in other cases, for instance \cite{benedetti2015functional,Lahoche:2024gal,finotello2026datafieldtheorytheory,Natta:2024nke}. Asymptotically in the IR, when $\rho(p)\sim \sqrt{p}$, the dimension tends to the value $1/2$, compatible with the correspondence with an ordinary Euclidean theory in dimension $D=3$ discussed previously. Figure \ref{figdim} shows the behavior of the canonical dimension for the quartic coupling. 

\begin{remark}
Although the focus here is on a quartic model, the asymptotic behavior of the distribution $\rho(p)$ induces a connection with the Pisarski fixed point for sextic models. Ordinarily, the existence of this fixed point is linked to dimensional regularization; the dependence of the canonical dimension on the scale $\mu$ plays an analogous role here, see \cite{Lahoche_2024PFP}.
\end{remark}

This dependence of the canonical dimension on the scale has an immediate consequence: there can be no global fixed points. This concept must be replaced here by that of a \textit{fixed trajectory}, a parametrized curve along which the beta functions vanish (see also \cite{Lahoche:2024gal}). These trajectories terminate in the deep IR, when the distribution begins to follow a power law and the canonical dimension becomes constant. The endpoint of these fixed trajectories defines the concept of an \textit{asymptotic fixed point}. 

Here, this asymptotic fixed point is given by the zeros of the asymptotic $\beta$-functions,
\begin{align}
\beta_m=-\bar{m}-\frac{\bar{u}}{1+\bar{m}}\,,\qquad \beta_u=-\frac{1}{2} \bar{u}+\frac{\bar{u}^2}{(1+\bar{m})^2}\,.
\end{align}
These equations admit an interacting fixed point (in addition to the Gaussian one), for the values:
\begin{equation}
(\bar{m}_*,\bar{u}_*)\to \left\{-\frac{1}{3},\frac{2}{9}\right\}\,,
\end{equation}
with the critical exponents:
\begin{equation}
\left\{\theta_1=\frac{\sqrt{3}}{2},\theta_2=-\frac{\sqrt{3}}{2}\right\}\,.
\end{equation}
{The interpretation of these exponents relies on the observation that, in the deep IR limit near the fixed point, the scaling behavior along the relevant direction is essentially governed by the dimensionless parameter $\bar{h}$ evaluated along the transition line. This parameter is related to the dimensionful coupling $h$ (analogous to the reduced temperature) via $\bar{h}:=h \mu^{-\theta_1}$. Since the deep-IR flow takes place near the soft spectral edge where $\rho(p) \sim \sqrt{p}$ (in agreement with the Tracy--Widom edge scaling recalled in Theorem \ref{thBBP}), the number of eigenvalues below the scale $\mu$ scales as $N \int_0^\mu \rho(p)\,\dd p \sim N \mu^{3/2}$, fixing the natural edge scaling $\mu \sim N^{-\ell}$ with $\ell = 2/3$. We therefore obtain $\bar{h} \sim h N^{\ell \theta_1}$. For the quartic truncation, this yields $\ell \theta_1 = \frac{2}{3}\frac{\sqrt{3}}{2} = \frac{1}{\sqrt{3}} \approx 0.58$, which compares qualitatively with the exact double-scaling exponent $2/3 \approx 0.67$ obtained in \eqref{Z_Ndoublescal} (this comparison must however be taken with care: as shown in Section \ref{nonperturbative}, this fixed point governs the BBP condensation boundary rather than the criticality of the original $O(N)$ model, and its exact exponent $\theta_0=1/2$ gives $\ell\theta_0=1/3$). At the same time, in the language of three-dimensional statistical field theory where $p$ plays the role of $\vec{p}\,^2$ and $\theta_1 = (2\nu)^{-1}$, the exact large-$N$ $O(N)$ Wilson--Fisher exponent ($\nu = 1$) corresponds to $\theta_1 = 0.5$, as further discussed in Remark \ref{rem:sign_vec} and Appendix \ref{App1}.}

\begin{remark}
Note that this construction naturally gives rise to an asymptotic just-renormalizable sector.
\end{remark}

\begin{remark}
Denoting the sextic coupling as $v$, normalized as $\frac{v}{6N^2}(\Phi\cdot\Phi)^3$ (see \eqref{exact_vk}) with $v=:\mu\rho^{-2}(\mu)\bar v$, the resulting flow equations including it are (the one for $\bar{m}$ being unchanged, and the sextic coupling having dimension zero, as expected):
\begin{align}
\beta_u = -\frac{1}{2}\bar{u} + \frac{\bar{u}^2}{(1+\bar{m})^2} - \frac{2\bar{v}}{1+\bar{m}}\,,\qquad
\beta_v = \frac{3\bar{u}\bar{v}}{(1+\bar{m})^2} - \frac{\bar{u}^3}{(1+\bar{m})^3}\,.
\end{align}
The resulting fixed point gives $\theta_1\approx 0.68$ (i.e., $\theta_1$ decreases from $\sqrt{3}/2\approx 0.87$ to $0.68$, moving closer to the exact three-dimensional large-$N$ $O(N)$ value $\theta_0 = 0.5$; correspondingly, $\ell\theta_1\approx 0.45$ moves \emph{away} from the double-scaling exponent $2/3$), for the coupling values $\{\bar{m}\approx -0.6,\bar{u}\approx 0.24,\bar{v}\approx 0.048\}$.
\end{remark}

{
\begin{remark}\label{rem:sign_vec}
Note that for $\beta \in (0, 1/2)$, the critical point $g_c = -1/4$ of the original quartic $O(N)$ model maps via \eqref{defgprime} to a negative bare coupling $u = (g_c + \beta^2)/\beta^2=1-1/(4\beta^2) < 0$ and a positive bare mass $m = \beta^{-1} - 2 > 0$ for the residual field $\Phi$, whereas the asymptotic Wilson--Fisher fixed point $(\bar{m}_*, \bar{u}_*) = (-1/3, 2/9)$ (and its sextic refinement above) lies in the confining quadrant $\bar{u}_* > 0$, $\bar{m}_* < 0$ for $\Phi$. Physically, one expects that this fixed point governs the continuous symmetry-breaking transition of the associated quenched $p = 2$ soft-spin model in effective dimension $D = 3$ \cite{de2006random,lahoche2023functional}, whose exact large-$N$ critical exponent is $\theta_1 = (2\nu)^{-1} = 0.5$ (i.e. corresponding to $\nu = 1$, see Appendix \ref{App1}). Within the perturbative quartic truncation, the axis $\bar{u} = 0$ is invariant under the flow, so a bare trajectory starting from $u < 0$ cannot reach the $\bar{u} > 0$ fixed point. Our expectation at this point is that the non-confining instability at $u < 0$ therefore lies beyond a simple polynomial vertex expansion around $\Phi = 0$ and requires a non-perturbative, non-polynomial treatment of the effective potential, which is left for the next section.
\end{remark}
}

\section{Non-perturbative analysis}\label{nonperturbative}

The perturbative Wilsonian flow analyzed in Section \ref{renormalization} exhibits an asymptotic interacting fixed point in the quadrant $(\bar{m}_* < 0, \bar{u}_* > 0)$, whose relevant critical exponent decreases from $\theta_1 = \sqrt{3}/2 \approx 0.87$ at quartic order to $\theta_1 \approx 0.68$ when the sextic coupling is included. Two natural questions are in order from this perturbative analysis: first, what is the exact non-perturbative status of this IR fixed point once all higher-order couplings generated along the flow are taken into account; and second, how does the relational background formalism capture the criticality ($g_c = -1/4$, $\gamma = 1/2$) of the original $O(N)$ vector model?

In this section, we address both questions within the non-perturbative Functional Renormalization Group (FRG) framework based on the Wetterich equation. For the $O(N)$ vector model, the Local Potential Approximation (LPA) turns out to be exact in the large-$N$ limit at all scales, from the UV cutoff $\Lambda = 4$ down to the deep IR ($k \to 0$). Integrating the resulting flow equation analytically shows that the effective theory \eqref{classicalaction} actually accommodates two distinct critical regimes, depending on the sign of the remnant quartic coupling $u$: a massless Wilson--Fisher transition in the confining remnant sector ($u > 0$), which corresponds to the BBP boundary $\beta \rho_0 = 1$ of the quenched $p=2$ model, and a finite-mass shock singularity in the non-confining remnant sector ($u < 0$), which reproduces the exact criticality of the original $O(N)$ vector model.

\subsection{Wetterich equation and exactness of the LPA at large $N$}

Let us begin by recalling explicitly how the bare parameters $(m,u)$ of the effective action \eqref{classicalaction} are related to the parameters $(g,\beta)$ of the partial Hubbard--Stratonovich decomposition \eqref{quartic2}--\eqref{defgprime}. Starting from the quenched partition function \eqref{modeleff} in the basis where $\sigma$ is diagonal, with shifted eigenvalues $p_i := \lambda_i + 2 \in [0,4]$ ($i=1,\dots,N$), the field rescaling $\phi_i \to \beta^{-1/2}\phi_i$ together with the definition $\Phi_i \equiv \Phi(p_i) := N \phi_i$ puts the bare action into the discrete form:
\begin{equation}
\mathcal{S}[\Phi] = \frac{1}{2N} \sum_{i=1}^N (p_i + m)\, \Phi_i^2 + \frac{u}{4N} \left(\frac{1}{N} \sum_{i=1}^N \Phi_i^2\right)^2\,,\label{discrete_action}
\end{equation}
where the bare mass $m$ and bare quartic coupling $u$ are given by:
\begin{equation}
\boxed{m := \beta^{-1} - 2\,, \qquad u := \frac{g^\prime}{\beta^2} = 1 + \frac{g}{\beta^2}\,.}\label{exact_bare_params}
\end{equation}
Note that keeping track of the rescaling factor $\beta^{-2}$ in $u = g^\prime/\beta^2$ does not modify the perturbative flow equations of Section \ref{renormalization}, but it is required for the quantitative identification of $g_c = -1/4$ below. It will be convenient to introduce the $O(N)$-invariant scalar composite:
\begin{equation}
\chi := \frac{1}{2N} (\Phi \cdot \Phi) = \frac{1}{2N^2} \sum_{i=1}^N \Phi_i^2\,,\label{def_chi}
\end{equation}
which is related to the original radial variable $\rho = \phi^2$ (before the $\beta^{-1/2}$ rescaling) through $\chi = \frac{1}{2}\beta \rho$. In terms of $\chi$, the bare action \eqref{discrete_action} takes the simple form:
\begin{equation}
\mathcal{S}[\Phi] = \frac{1}{2N} \sum_{i=1}^N p_i\, \Phi_i^2 + N U_\Lambda(\chi)\,, \qquad U_\Lambda(\chi) := m\chi + u\chi^2\,.\label{bare_pot}
\end{equation}

To construct a non-perturbative RG flow with respect to the relational momentum $p_i$, we add a scale-dependent IR regulator term to the bare action, quadratic in the field and diagonal in the eigenbasis of $\sigma$:
\begin{equation}
\Delta \mathcal{S}_k[\Phi] := \frac{1}{2N} \sum_{i=1}^N R_k(p_i)\,\Phi_i^2 = \frac{1}{2} \sum_{i,j=1}^N \mathcal{R}_{k,ij}\,\Phi_i \Phi_j\,, \qquad \mathcal{R}_{k,ij} := \frac{1}{N} R_k(p_i)\,\delta_{ij}\,,
\end{equation}
where $k \in [0,\Lambda]$ denotes the running IR scale (with $\Lambda = 4$). The regulator function $R_k(p)$ is chosen so as to interpolate between the UV and IR regimes:
\begin{itemize}
    \item $R_{k\to 0}(p) = 0$ for all $p \in [0,4]$, so that all fluctuations are integrated out in the IR limit;
    \item $R_{k\to \Lambda}(p) \to \infty$, freezing the fluctuations in the UV so that the effective average action $\Gamma_k$ approaches the bare action $\mathcal{S}$;
    \item $R_k(p) > 0$ for $p < k$ (endowing IR modes with an effective mass of order $k$) and $R_k(p) \to 0$ for $p > k$ (leaving UV modes unaffected).
\end{itemize}
Introducing the scale-dependent generating functional $W_k[J] := \ln \int \prod_{i=1}^N \dd \Phi_i\, e^{-\mathcal{S}[\Phi] - \Delta \mathcal{S}_k[\Phi] + \sum_{i=1}^N J_i \Phi_i}$ and the effective average action via the modified Legendre transform $\Gamma_k[\Phi] := \sup_J \big(\sum_{i=1}^N J_i \Phi_i - W_k[J]\big) - \Delta \mathcal{S}_k[\Phi]$ (where $\Phi_i$ now denotes the classical field $\langle \Phi_i \rangle$), $\Gamma_k[\Phi]$ obeys the exact Wetterich equation:
\begin{equation}
\partial_t \Gamma_k[\Phi] = \frac{1}{2}\,\Tr\!\left[\left(\Gamma_k^{(2)}[\Phi] + \mathcal{R}_k\right)^{-1} \partial_t \mathcal{R}_k\right]\,, \qquad t := \ln k\,,\label{wetterich_vec}
\end{equation}
where $\Gamma_{k,ij}^{(2)}[\Phi] := \frac{\partial^2 \Gamma_k[\Phi]}{\partial \Phi_i \partial \Phi_j}$ is the second functional derivative of $\Gamma_k$.

Let us now consider the Local Potential Approximation (LPA), in which the kinetic term is unrenormalized (no anomalous dimension, $Z_k = 1$) while the effective potential $U_k(\chi)$ is left as an arbitrary function of $\chi$, thus retaining all local interactions of arbitrary order ($u_4, u_6, u_8, \dots$):
\begin{equation}
\Gamma_k[\Phi] = \frac{1}{2N} \sum_{i=1}^N p_i\,\Phi_i^2 + N U_k(\chi)\,.\label{LPA_ansatz}
\end{equation}
Since $\frac{\partial \chi}{\partial \Phi_i} = \frac{1}{N^2}\Phi_i$ and $\frac{\partial^2 \chi}{\partial \Phi_i \partial \Phi_j} = \frac{1}{N^2}\delta_{ij}$, the first and second derivatives of \eqref{LPA_ansatz} read:
\begin{align}
\frac{\partial \Gamma_k[\Phi]}{\partial \Phi_i} &= \frac{1}{N}\big(p_i + U^\prime_k(\chi)\big)\Phi_i\,,\\
\Gamma_{k,ij}^{(2)}[\Phi] + \mathcal{R}_{k,ij} &= \frac{1}{N}\!\left[D_k(p_i;\chi)\,\delta_{ij} + \frac{1}{N^2} U^{\prime\prime}_k(\chi)\,\Phi_i \Phi_j\right]\,,
\end{align}
where we have defined the regularized inverse transverse propagator:
\begin{equation}
D_k(p;\chi) := p + R_k(p) + U^\prime_k(\chi)\,.
\end{equation}
Because the regularized Hessian is the sum of a diagonal matrix and a rank-$1$ matrix, its inverse can be computed explicitly at any finite $N$ using the well known Sherman--Morrison identity (see for instance \cite{potters2020first}):
\begin{equation}
\left(\Gamma_k^{(2)}[\Phi] + \mathcal{R}_k\right)^{-1}_{ij} = \frac{N}{D_k(p_i;\chi)}\,\delta_{ij} - \frac{1}{N}\,\frac{U^{\prime\prime}_k(\chi)}{1 + \frac{U^{\prime\prime}_k(\chi)}{N^2}\sum_{l=1}^N \frac{\Phi_l^2}{D_k(p_l;\chi)}}\,\frac{\Phi_i \Phi_j}{D_k(p_i;\chi) D_k(p_j;\chi)}\,.
\end{equation}
Multiplying by $\partial_t \mathcal{R}_{k,ij} = \frac{1}{N}\dot{R}_k(p_i)\delta_{ij}$ (with $\dot{R}_k := \partial_t R_k$) and taking the trace, the Wetterich equation projected onto the LPA ansatz becomes, at finite $N$:
\begin{equation}
\partial_t \Gamma_k[\Phi] = \frac{1}{2} \sum_{i=1}^N \frac{\dot{R}_k(p_i)}{D_k(p_i;\chi)} - \frac{1}{2N^2}\,\frac{U^{\prime\prime}_k(\chi)}{1 + \frac{U^{\prime\prime}_k(\chi)}{N^2}\sum_{l=1}^N \frac{\Phi_l^2}{D_k(p_l;\chi)}} \sum_{i=1}^N \frac{\dot{R}_k(p_i)\,\Phi_i^2}{D_k(p_i;\chi)^2}\,.\label{wetterich_finiteN}
\end{equation}
Hence, because $\chi = \frac{1}{2N^2}\sum_{i=1}^N \Phi_i^2 = \mathcal{O}(1)$ (see also \cite{delamotte2012introduction}):
\begin{itemize}
    \item The first term, associated with the $N-1$ transverse modes, is of order $\mathcal{O}(N)$ and depends on the field $\Phi$ only through the invariant $\chi$.
    \item The second term, associated with the single longitudinal (radial) mode, is of order $\mathcal{O}(1)$ — that is, suppressed by a factor $1/N$ relative to the leading contribution as long as $U^{\prime\prime}_k(\chi)$ remains finite.
\end{itemize}
Note that, in the non-vacuum expansion, it is precisely this second term that would generate wave-function renormalization ($\eta_k \neq 0$) and momentum-dependent (derivative) interactions. In the limit $N \to \infty$, it decouples at every scale $k \in [0,\Lambda]$, including in the UV region near $k = 4$: the LPA ansatz \eqref{LPA_ansatz} is therefore closed and exact at leading order in $1/N$. Dividing \eqref{wetterich_finiteN} by $N$ and using the continuous formulation, we obtain the exact large-$N$ flow equation for the effective potential:
\begin{equation}
\boxed{\partial_t U_k(\chi) = \frac{1}{2} \int_0^4 \dd p\,\rho(p)\,\frac{\dot{R}_k(p)}{p + R_k(p) + U^\prime_k(\chi)}\,, \qquad \rho(p) = \frac{\sqrt{p(4-p)}}{2\pi}\,.}\label{exact_LPA_flow}
\end{equation}

\subsection{Integration using the method of characteristics}

Equation \eqref{exact_LPA_flow} can be integrated analytically for any regulator $R_k(p)$ using the standard method of characteristics. Let us define the field-dependent effective mass:
\begin{equation}
W_k(\chi) := U^\prime_k(\chi)\,,
\end{equation}
and the regulated one-loop integral:
\begin{equation}
I_k(W) := \frac{1}{2} \int_0^4 \dd p\,\frac{\rho(p)}{p + R_k(p) + W} =: \frac{1}{2} F_k(W)\,,\label{def_Ik}
\end{equation}
whose scale derivative at fixed $W$ reads:
\begin{equation}
\partial_t I_k(W) = -\frac{1}{2} \int_0^4 \dd p\,\frac{\rho(p)\,\dot{R}_k(p)}{\big(p + R_k(p) + W\big)^2}\,.
\end{equation}
Differentiating the flow equation \eqref{exact_LPA_flow} with respect to $\chi$ gives a first-order quasi-linear partial differential equation of Burgers type for $W_k(\chi)$ (see Appendix \ref{App2}):
\begin{equation}
\partial_t W_k(\chi) = \Big(\partial_t I_k(W)\Big\vert_{W=W_k(\chi)}\Big)\,\partial_\chi W_k(\chi)\,.\label{burgers_eq}
\end{equation}
Assuming locally that $W_k(\chi)$ is invertible, we introduce the inverse function $\chi = \chi_k(W)$, defined by $W_k(\chi_k(W)) \equiv W$. Differentiating this relation with respect to $t$ at fixed $W$ yields:
\begin{equation}
(\partial_t W_k)(\chi_k(W)) + (\partial_\chi W_k)(\chi_k(W))\,\partial_t \chi_k(W) = 0\,.
\end{equation}
Substituting \eqref{burgers_eq} into this relation eliminates $\partial_\chi W_k$, and reduces the flow equation along lines of constant $W$ to:
\begin{equation}
\partial_t \chi_k(W) = -\partial_t I_k(W)\,.
\end{equation}
Integrating from the UV cutoff $k = \Lambda$ (where $R_\Lambda(p) \to \infty$, so that $I_\Lambda(W) = 0$) down to an arbitrary scale $k$, we find:
\begin{equation}
\chi_k(W) = \chi_\Lambda(W) - I_k(W)\,.
\end{equation}
From the bare potential \eqref{bare_pot}, we have $W_\Lambda(\chi) = m + 2u\chi$, and therefore $\chi_\Lambda(W) = \frac{W - m}{2u}$. Substituting this initial condition back into the integrated equation yields the exact implicit solution at any scale $k \in [0,\Lambda]$:
\begin{equation}
\boxed{W_k(\chi) = m + 2u\chi + u \int_0^4 \dd p\,\frac{\rho(p)}{p + R_k(p) + W_k(\chi)} = m + 2u\chi + u F_k\big(W_k(\chi)\big)\,.}\label{exact_characteristic_sol}
\end{equation}

Now, defining the running mass $m_k := W_k(0)$, the running quartic coupling $u_k := \frac{1}{2}U^{\prime\prime}_k(0) = \frac{1}{2}W^\prime_k(0)$, and the running sextic coupling $v_k := \frac{1}{8}U^{\prime\prime\prime}_k(0) = \frac{1}{8}W^{\prime\prime}_k(0)$ (corresponding to $\frac{v}{6N^2}(\Phi\cdot\Phi)^3 = \frac{4}{3}N v\chi^3$), and differentiating \eqref{exact_characteristic_sol} twice with respect to $\chi$, we obtain:
\begin{align}
u_k(\chi) := \frac{1}{2}W^\prime_k(\chi) &= \frac{u}{1 - u F^\prime_k\big(W_k(\chi)\big)} = \frac{u}{1 + u \int_0^4 \dd p\,\frac{\rho(p)}{\big(p + R_k(p) + W_k(\chi)\big)^2}}\,,\label{exact_uk}\\
v_k(\chi) := \frac{1}{8}W^{\prime\prime}_k(\chi) &= \frac{1}{2} u_k(\chi)^3 F^{\prime\prime}_k\big(W_k(\chi)\big) = u_k(\chi)^3 \int_0^4 \dd p\,\frac{\rho(p)}{\big(p + R_k(p) + W_k(\chi)\big)^3}\,.\label{exact_vk}
\end{align}

In the physical IR limit $k \to 0$, the regulator vanishes ($R_0(p) = 0$) and $F_0(W) \equiv F(W)$ reduces to the exact Stieltjes transform of the shifted Wigner distribution $\rho(p) = \frac{\sqrt{p(4-p)}}{2\pi}$. Then, it obeys the algebraic identity:
\begin{equation}
F(W) + \frac{1}{F(W)} = W + 2\,.\label{alg_id_F}
\end{equation}
leading to:
\begin{equation}
F(W) = \frac{W + 2 - \sqrt{W(W+4)}}{2}\,, \qquad (W \geq 0)\,.\label{stieltjes_F}
\end{equation}
At the lower edge of the spectral support, $F(0) = 1$, and its asymptotic expansion for $W \to 0^+$ exhibits a square-root branch point:
\begin{equation}
F(W) = 1 - \sqrt{W} + \frac{W}{2} + \mathcal{O}(W^{3/2})\,,
\end{equation}
while $F(W) \sim W^{-1}$ as $W \to \infty$. Differentiating \eqref{stieltjes_F} and \eqref{alg_id_F} yields the explicit expressions for the first two derivatives for $W > 0$:
\begin{align}
F^\prime(W) &= \frac{1}{2}\!\left(1 - \frac{W+2}{\sqrt{W(W+4)}}\right) = -\frac{F(W)^2}{1 - F(W)^2} < 0\,,\label{Fp_exact}\\
F^{\prime\prime}(W) &= \frac{2}{\big(W(W+4)\big)^{3/2}} > 0\,.\label{Fpp_exact}
\end{align}
In particular, note that $F^\prime(W)$ is strictly negative for all $W > 0$ and diverges as $F^\prime(W) \sim -\frac{1}{2\sqrt{W}}$ as $W \to 0^+$.

Evaluating \eqref{exact_characteristic_sol} and \eqref{exact_uk} at $k=0$ and $\chi=0$, the exact renormalized mass $M := W_0(0)$ and quartic coupling $u_0(0)$ satisfy:
\begin{equation}
\boxed{M = m + u F(M)\,, \qquad u_0(0) = \frac{u}{1 - u F^\prime(M)}\,.}\label{k0_gap_and_u}
\end{equation}
Note that \eqref{k0_gap_and_u} results from integrating the flow over the entire spectral support $p \in [0,4]$, and therefore fully incorporates the UV regime near $k = 4$ where $\dim(u)$ diverges. In particular, if $m > 0$ and $u > 0$ in the UV, then $M > m > 0$ at $k = 0$: the UV fluctuations shift the effective mass upward and therefore cannot flip its sign. From these observations, we have to analyze the two sectors $u > 0$ and $u < 0$ separately.

\subsection{Confining remnant sector ($u > 0$)}

We first consider the regime $u > 0$, in which the bare quartic potential of the remnant field $\Phi$ is confining. Because $F^\prime(W) < 0$ for all $W > 0$, the denominator $1 - u F^\prime(M) > 1$ in \eqref{k0_gap_and_u} never vanishes, and the only non-analyticity occurs when the physical mass $M$ vanishes, $M \to 0^+$. From \eqref{k0_gap_and_u} and $F(0)=1$, this defines a critical bare mass:
\begin{equation}
\boxed{m_c = -u F(0) = -u < 0\,.}
\end{equation}
Expressing this critical condition in terms of the original parameters $(g,\beta)$ via \eqref{exact_bare_params}, we therefore obtain:
\begin{equation}
m = m_c \iff \beta^{-1} - 2 = -\!\left(1 + \frac{g}{\beta^2}\right) \iff g\left(\frac{1}{\beta}\right)^2 + \frac{1}{\beta} - 1 = 0\,.
\end{equation}
Comparing this relation with the saddle-point equation \eqref{condition_sadle_quart} of the $O(N)$ model ($g\rho_0^2 + \rho_0 - 1 = 0$), we see that $m = m_c$ coincides exactly with $\rho_0 = \beta^{-1}$, that is, with the BBP boundary $\beta\rho_0 = 1$ discussed above. 

Near this transition, let $h := m - m_c = m + u$ denote the reduced temperature (see Appendix \ref{App1}). Subtracting $m_c = -u F(0)$ from the gap equation \eqref{k0_gap_and_u} yields:
\begin{equation}
h = M + u\big(F(0) - F(M)\big) = M + u M \int_0^4 \dd p\,\frac{\rho(p)}{p(p+M)}\,.
\end{equation}
For $M \ll 1$, the integral is dominated by the IR regime, such that $\rho(p) \approx \frac{1}{\pi}\sqrt{p}$, so that $\int_0^4 \dd p\,\frac{\rho(p)}{p(p+M)} \approx \frac{1}{\pi}\int_0^\infty \frac{\dd p}{\sqrt{p}(p+M)} = \frac{1}{\sqrt{M}}$ (which also follows directly from the small-$M$ expansion $F(M) = 1 - \sqrt{M} + \mathcal{O}(M)$). Hence, as $h \to 0^+$:
\begin{equation}
h \approx u\sqrt{M} \implies M \sim \left(\frac{h}{u}\right)^2\,, \qquad u_0(0) = \frac{u}{1 - u F^\prime(M)} \approx 2\sqrt{M} \sim \frac{2h}{u} \to 0^+\,.\label{exact_WF_scaling}
\end{equation}
Since the physical mass $M$ sets the IR scale of the generalized momentum ($\mu \sim M$, corresponding to $\xi^{-2}$ in a $3$-dimensional Euclidean field theory), we deduce the correlation-length exponent $\nu = 1$ and the exact relevant critical exponent associated with the generalized momentum scale:
\begin{equation}
h \sim M^{1/2} \implies \theta_0 = \frac{1}{2\nu} = \frac{1}{2}\,.
\end{equation}

\paragraph{IR fixed point with Litim's regulator.}
Let us now check directly how this transition emerges from the asymptotic, IR, fixed point of the Wetterich flow \eqref{exact_LPA_flow}. Consider the standard Litim regulator, which is well suited to the deep IR (see \cite{Lahoche_2026_background}):
\begin{equation}
R_k(p) = (k - p)\,\theta(k - p)\,, \qquad \dot{R}_k(p) = k\,\theta(k - p)\,,\label{litim_reg}
\end{equation}
the flow equation \eqref{exact_LPA_flow} takes the form:
\begin{equation}
\partial_t U_k(\chi) = \frac{1}{2}\,\frac{k}{k + U^\prime_k(\chi)} \int_0^k \dd p\,\rho(p)\,.
\end{equation}
In the deep IR ($k \ll 1$), we have $\rho(p) \approx \frac{1}{\pi}\sqrt{p}$ and $\int_0^k \dd p\,\rho(p) \approx \frac{2}{3\pi} k^{3/2}$, giving:
\begin{equation}
\partial_t U_k(\chi) = \frac{1}{3\pi}\,\frac{k^{3/2}}{1 + k^{-1}U^\prime_k(\chi)}\,.
\end{equation}
To obtain an autonomous system in the IR, we introduce the dimensionless potential and field (the $3\pi$ factor is conventional):
\begin{equation}
\tilde{U}_k(\tilde{\chi}) := 3\pi\, k^{-3/2} U_k(\chi)\,, \qquad \tilde{\chi} := 3\pi\, k^{-1/2} \chi\,,
\end{equation}
so that the dimensionless mass function $\bar{W}_k(\tilde{\chi}) := \tilde{U}^\prime_k(\tilde{\chi}) = k^{-1}U^\prime_k(\chi)$ preserves the asymptotic canonical dimensions $\dim(m) = 1$, $\dim(u) = 1/2$, and $\dim(v) = 0$ obtained in Section \ref{renormalization}. The dimensionless flow equation reads:
\begin{equation}
\partial_t \tilde{U}_k(\tilde{\chi}) = -\frac{3}{2}\tilde{U}_k(\tilde{\chi}) + \frac{1}{2}\tilde{\chi}\,\tilde{U}^\prime_k(\tilde{\chi}) + \frac{1}{1 + \tilde{U}^\prime_k(\tilde{\chi})}\,.
\end{equation}
Differentiating with respect to $\tilde{\chi}$ yields the autonomous equation for $\bar{W}_k(\tilde{\chi})$:
\begin{equation}
\partial_t \bar{W}_k(\tilde{\chi}) = -\bar{W}_k(\tilde{\chi}) + \left(\frac{1}{2}\tilde{\chi} - \frac{1}{\big(1 + \bar{W}_k(\tilde{\chi})\big)^2}\right)\bar{W}^\prime_k(\tilde{\chi})\,,
\end{equation}
which, in terms of the inverse function $\tilde{\chi}_k(\bar{W})$ (using $\partial_t \bar{W}_k = -\bar{W}^\prime_k\,\partial_t \tilde{\chi}_k$), becomes linear in $\tilde{\chi}_k$:
\begin{equation}
\partial_t \tilde{\chi}_k(\bar{W}) = \bar{W}\,\tilde{\chi}^\prime_k(\bar{W}) - \frac{1}{2}\tilde{\chi}_k(\bar{W}) + \frac{1}{(1 + \bar{W})^2}\,.\label{dimless_chi_flow}
\end{equation}

At a fixed point ($\partial_t \tilde{\chi}_* = 0$), equation \eqref{dimless_chi_flow} reduces to a first-order linear differential equation:
\begin{equation}
\tilde{\chi}^\prime_*(\bar{W}) - \frac{1}{2\bar{W}}\tilde{\chi}_*(\bar{W}) = -\frac{1}{\bar{W}(1 + \bar{W})^2}\,.
\end{equation}
Multiplying by the integrating factor $\bar{W}^{-1/2}$ (for $\bar{W} > 0$) and integrating via the substitution $\bar{W} = x^2$, we find the general solution:
\begin{equation}
\tilde{\chi}_*(\bar{W}) = 2 + \frac{\bar{W}}{1 + \bar{W}} + 3\sqrt{\bar{W}}\arctan\!\sqrt{\bar{W}} + C\sqrt{\bar{W}}\,,
\end{equation}
where $C$ is an integration constant. Requiring the fixed-point potential to be analytic across its minimum $\bar{W} = 0$ fixes $C = 0$, and finally:
\begin{equation}
\boxed{\tilde{\chi}_*(\bar{W}) = 2 + \frac{\bar{W}}{1 + \bar{W}} + 3\sqrt{\bar{W}}\arctan\!\sqrt{\bar{W}}\,, \qquad (\bar{W} > -1)\,,}\label{exact_FP_potential}
\end{equation}
which extends analytically to $-1 < \bar{W} < 0$ because of the standard identity $\sqrt{\bar{W}}\arctan\sqrt{\bar{W}} = -\sqrt{-\bar{W}}\operatorname{arctanh}\sqrt{-\bar{W}}$. At the minimum $\bar{W} = 0$, the dimensionless condensate sits at $\tilde{\chi}_*(0) = 2$, while at the origin $\tilde{\chi} = 0$, the condition $\tilde{\chi}_*(\bar{m}_*) = 0$ gives a unique negative dimensionless mass $\bar{m}_* \approx -0.388$ and a positive quartic coupling $\bar{u}_* > 0$. Finally, observing that, at finite $k > 0$, the Litim regulator shifts the spectral singularity from $\bar{W} = 0$ to $\bar{W} = -1$, so $\bar{m}_* \approx -0.388$ remains inside the analytic domain of the flow.

The critical exponents around \eqref{exact_FP_potential} are obtained by linearizing \eqref{dimless_chi_flow} around the fixed point, writing $\tilde{\chi}_k(\bar{W}) = \tilde{\chi}_*(\bar{W}) + \epsilon\, e^{-\theta t} f(\bar{W})$ (with $\theta > 0$ corresponding to a relevant direction growing toward the IR, $t \to -\infty$). Because \eqref{dimless_chi_flow} is linear in $\tilde{\chi}_k$, the eigenfunction $f(\bar{W})$ satisfies:
\begin{equation}
\bar{W}\, f^\prime(\bar{W}) = \left(\frac{1}{2} - \theta\right) f(\bar{W}) \implies f(\bar{W}) = A\,\bar{W}^{\frac{1}{2} - \theta}\,.
\end{equation}
Requiring analyticity of the perturbation at the minimum $\bar{W} = 0$ quantizes the exponent as $\frac{1}{2} - \theta = n \in \mathbb{N}$, which yields the exact spectrum of critical exponents:
\begin{equation}
\boxed{\theta_n = \frac{1}{2} - n\,.}\label{exact_spectrum}
\end{equation}
There is therefore a single relevant direction, with exact critical exponent $\theta_0 = 1/2$, while all other directions ($n \geq 1$) are irrelevant. This explains the trend observed in Section \ref{renormalization}, where the perturbative estimate decreased from $\theta_1 = \sqrt{3}/2 \approx 0.87$ at quartic order to $\theta_1 \approx 0.68$ at sextic order, moving toward the (exact) LPA value $\theta_0 = 0.5$.

\begin{remark}
This non-perturbative solution clarifies the physical meaning of the asymptotic fixed point obtained in Section \ref{renormalization}. For $g > 0$, the original zero-dimensional $O(N)$ vector model \eqref{modelquartic} has an analytic free energy for any quadratic coupling $m_0 \in \mathbb{R}$ and exhibits no symmetry-breaking phase transition. The Wilson--Fisher fixed point \eqref{exact_FP_potential} (belonging to the 3D large-$N$ $O(N)$ / spherical universality class, with $\nu = 1$ and $\theta_0 = 1/2$) governs the genuine second-order condensation transition of the \textit{quenched} $p = 2$ soft-spin model \eqref{modeleff}, in which the spectrum of $\sigma$ is rigidly frozen onto the Wigner semicircle but beyond the sector of the phase space where the physics of this model coincides with the $O(N)$ model. Indeed, from the viewpoint of the original \textit{annealed} $O(N)$ model, reaching this critical surface ($m = m_c \iff \beta \rho_0 = 1$) does not correspond to a singularity of $F_N(g)$, but rather to the BBP threshold where an isolated eigenvalue detaches from the bulk and the quenched--annealed equivalence breaks down.
\end{remark}

\subsection{Non-confining remnant sector ($u < 0$)}

Let us now turn to the sector $u < 0$, which is where the genuine criticality of the original $O(N)$ vector model lies. Indeed, recall that, for any splitting parameter chosen inside the BBP validity domain,
\begin{equation}
0 < \beta < \frac{1}{2}\,,
\end{equation}
the bare mass $m = \beta^{-1} - 2$ is strictly positive. As the coupling $g$ is lowered from $0$ toward $g_c = -1/4$, the remnant quartic coupling $u = 1 + g/\beta^2$ first remains positive in the metastable band $-\beta^2 < g < 0$, vanishes at $g = -\beta^2$, and becomes strictly negative before $g_c = -1/4$ is reached (since $\beta^2 < 1/4$).

Let us first verify that the exact $k=0$ gap equation \eqref{k0_gap_and_u} reproduces the large-$N$ saddle point of the $O(N)$ model for any $\beta \in (0,1/2)$. For $\chi=0$, substituting $W = F(W) + F(W)^{-1} - 2$ from \eqref{alg_id_F} together with \eqref{exact_bare_params} into $W - m = u F(W)$, we find:
\begin{equation}
F(W) + \frac{1}{F(W)} - \frac{1}{\beta} = \left(1 + \frac{g}{\beta^2}\right) F(W) \iff g\left(\frac{F(W)}{\beta}\right)^2 + \frac{F(W)}{\beta} - 1 = 0\,.
\end{equation}
Setting $\rho_0 := F(W)/\beta$, we recover again the saddle-point equation \eqref{condition_sadle_quart} of the original $O(N)$ model, with the BBP condition $\beta \rho_0 < 1$, because: $0 < F(W) < 1$, i.e., $W > 0$.

When $u < 0$, the denominator $1 - u F^\prime(W) = 1 - \vert u \vert\,\vert F^\prime(W) \vert$ in \eqref{k0_gap_and_u} can vanish at a finite, strictly positive mass $W_c > 0$. At this point, the characteristic curve $\chi_0(W)$ verifies $\partial_W \chi_0(W_c) = 0$ and the Burgers solution $W_0(\chi)$ encounters a shock singularity where the renormalized quartic coupling $u_0(0) = \frac{1}{2}U^{\prime\prime}_0(0)$ diverges to $-\infty$. Let us determine the exact location of this shock. Substituting \eqref{Fp_exact}, namely $F^\prime(W_c) = -\frac{F_c^2}{1 - F_c^2}$ with $F_c := F(W_c) = \beta \rho_0$, into the shock condition $1 - u_c F^\prime(W_c) = 0$ gives:
\begin{equation}
1 + u_c\,\frac{F_c^2}{1 - F_c^2} = 0 \iff 1 - F_c^2 + \left(1 + \frac{g_c}{\beta^2}\right) F_c^2 = 0 \iff g_c\left(\frac{F_c}{\beta}\right)^2 = -1\,.
\end{equation}
We recognize the exact criticality condition $g_c \rho_0^2 = -1$ of \eqref{critical_cond}. Combining it with $g_c \rho_0^2 + \rho_0 - 1 = 0$ immediately yields:
\begin{equation}
\rho_0 = 2\,, \qquad g_c = -\frac{1}{4}\,,
\end{equation}
independently of the choice of $\beta \in (0,1/2)$. At the shock, we have $F_c = 2\beta \in (0,1)$, so that the critical effective mass $W_c$ and critical bare coupling $u_c$ read explicitly:
\begin{equation}
\boxed{W_c = F_c + \frac{1}{F_c} - 2 = \frac{(1 - 2\beta)^2}{2\beta} > 0\,, \qquad u_c = 1 - \frac{1}{4\beta^2} = -\frac{1 - 4\beta^2}{4\beta^2} < 0\,.}\label{shock_values}
\end{equation}
Note again that, for every $\beta \in (0,1/2)$, the transition at $g_c = -1/4$ occurs at a strictly positive effective mass $W_c > 0$ for the remnant field $\Phi$ (and therefore away from the branch point $W = 0$). Because the gap $W_c$ remains open at criticality, the RG flow of the transverse modes effectively freezes for $k \ll W_c$ and does not flow to a massless asymptotic fixed point as $k \to 0$. Only in the degenerate boundary limit $\beta \to 1/2^-$ does $W_c \to 0^+$ coincide with $g \to g_c$, in which case we have $m \to 0$ and $u_c \to 0^-$ simultaneously, and the classical action $\mathcal{S}[\Phi]$ becomes purely Gaussian.

\paragraph{Critical scaling.}
To extract the critical behavior in the vicinity of the shock, let us expand the exact gap equation $W - m = u F(W)$ around $(u_c, W_c)$ by setting $u = u_c + \delta u$ (with $\delta u = \beta^{-2}(g - g_c) > 0$) and $W = W_c + \delta W$:
\begin{equation}
\delta W = \delta u\, F(W_c) + u_c F^\prime(W_c)\,\delta W + \frac{1}{2} u_c F^{\prime\prime}(W_c)\,(\delta W)^2 + \mathcal{O}(\delta u\,\delta W, \delta W^3)\,.
\end{equation}
Grouping the linear terms in $\delta W$ on the left-hand side gives:
\begin{equation}
\big(1 - u_c F^\prime(W_c)\big)\,\delta W = \delta u\, F(W_c) + \frac{1}{2} u_c F^{\prime\prime}(W_c)\,(\delta W)^2 + \mathcal{O}(\delta u\,\delta W, \delta W^3)\,.
\end{equation}
By the shock condition $1 - u_c F^\prime(W_c) = 0$, the left-hand side vanishes identically. Since $u_c = -\vert u_c \vert < 0$, we are left with the quadratic relation:
\begin{equation}
\frac{1}{2} \vert u_c \vert F^{\prime\prime}(W_c)\,(\delta W)^2 \approx \delta u\, F(W_c)\,.
\end{equation}
From \eqref{shock_values}, it is easy to check that $W_c(W_c + 4) = \frac{(1 - 4\beta^2)^2}{4\beta^2}$, which substituted into \eqref{Fpp_exact} gives $F^{\prime\prime}(W_c) = \frac{16\beta^3}{(1 - 4\beta^2)^3}$ and $\vert u_c \vert F^{\prime\prime}(W_c) = \frac{4\beta}{(1 - 4\beta^2)^2}$. Retaining the physical branch ($\delta W > 0$ for $g > g_c$, continuously connected to $g = 0$), we find:
\begin{equation}
\delta W = \sqrt{\frac{2 F(W_c)}{\vert u_c \vert F^{\prime\prime}(W_c)}\,\delta u} = \frac{1 - 4\beta^2}{\beta}\,\sqrt{g - g_c}\,,\label{deltaW_scaling}
\end{equation}
as well as the singular behavior of the renormalized quartic coupling:
\begin{equation}
1 - u F^\prime(W) \approx \vert u_c \vert F^{\prime\prime}(W_c)\,\delta W = \frac{4}{1 - 4\beta^2}\,\sqrt{g - g_c} \implies u_0(0) \approx -\frac{(1 - 4\beta^2)^2}{16\beta^2}\,(g - g_c)^{-1/2}\,.\label{u0_div}
\end{equation}
From $\delta F = F^\prime(W_c)\,\delta W = -4\beta\sqrt{g - g_c}$, we obtain $\rho_0 = F/\beta = 2 - 4\sqrt{g - g_c} + \mathcal{O}(g - g_c)$. Substituting this into the derivative of the free energy $\frac{\dd F_N}{\dd g} = \frac{1}{4}\rho_0^2 = 1 - 4\sqrt{g - g_c} + \mathcal{O}(g - g_c)$ directly reproduces the exact singular free energy and the branched-polymer entropic exponent:
\begin{equation}
F_{N,\text{sing}}(g) \sim (g - g_c)^{3/2} = (g - g_c)^{2 - \gamma}\,, \qquad \gamma = \frac{1}{2}\,.
\end{equation}

As a concluding remark, note that this result also indicates how the double-scaling regime emerges within the relational framework. Although the $N-1$ transverse modes retain a finite mass $W_c > 0$ at $g_c = -1/4$, the divergence of the longitudinal susceptibility $U^{\prime\prime}_0(0) = 2u_0(0) \sim -(g - g_c)^{-1/2}$ in \eqref{u0_div} signals that the single longitudinal (radial) mode becomes massless at the shock, compensating the $1/N$ suppression of the second term in the finite-$N$ Wetterich equation \eqref{wetterich_finiteN}. Moreover, the singular part of the total effective action scales as $N F_{N,\text{sing}}(g) \sim N(g - g_c)^{3/2}$, and becomes of order $\mathcal{O}(1)$ in the corresponding double-scaling window $g - g_c \sim N^{-2/3}$. 

\begin{claim}
The non-perturbative LPA flow of the relational effective action \eqref{classicalaction} splits into two distinct critical regimes:
\begin{enumerate}
    \item In the confining remnant sector ($u > 0$), the exact large-$N$ flow admits a massless asymptotic Wilson--Fisher fixed point \eqref{exact_FP_potential} with exact critical exponent spectrum $\theta_n = \frac{1}{2} - n$ ($n \geq 0$). This fixed point governs the BBP condensation boundary ($\beta\rho_0 = 1$) of the quenched $p = 2$ soft-spin model.
    \item In the non-confining remnant sector ($u < 0$), for any $\beta \in (0, 1/2)$, the exact large-$N$ flow develops a finite-mass shock singularity ($W_c > 0$) at $g_c = -1/4$, reproducing the exact branched-polymer entropic exponent $\gamma = 1/2$ and the double-scaling behavior $g - g_c \sim N^{-2/3}$ independently of $\beta$.
\end{enumerate}
\end{claim}

\section{Discussion and open issues}\label{conclusion}

In this pedagogical note, we have detailed the construction of a relational background field method for zero-dimensional $O(N)$ vector models. Using the BBP (Baik--Ben Arous--Péché) phase transition, we have shown how a partial Hubbard--Stratonovich decoupling allows an intrinsic notion of scale to emerge in the large-$N$ continuum limit. This mechanism avoids the ad hoc breaking of the native $O(N)$ symmetry that arises in standard Wilsonian partial-integration approaches. Moreover, because the RG scale is subordinated to the continuum limit, the inconsistency between the large-$N$ approximation and the deep IR limit is removed. At the perturbative level, the flow equations exhibit an asymptotic Wilson--Fisher-type fixed point in the confining sector of the residual field ($u > 0$). The exact non-perturbative LPA analysis of Section \ref{nonperturbative} then clarifies the full phase structure: this fixed point (with exact exponent $\theta_0 = 1/2$, i.e. $\nu = 1$) governs the BBP condensation boundary of the quenched $p=2$ model, whereas the criticality of the original $O(N)$ model ($g_c = -1/4$, $\gamma = 1/2$) is recovered in the non-confining sector ($u < 0$) as a finite-mass shock singularity whose longitudinal fluctuations single out the double-scaling window $g - g_c \sim N^{-2/3}$.

The most direct extension of this formalism is its application to higher-rank models: random matrix models and, ultimately, random tensor models and Group Field Theories. The case of random matrices has already been studied in \cite{Lahoche_2026_background}, where the reader will find an extensive discussion of the questions surveyed here. As noted in the text, vector models are protected against wave-function renormalization and non-trivial modifications of the Ward identities at leading order in $1/N$ \cite{Natta:2024nke}. It is worth emphasizing that these Ward identities reflect the preservation of the structural symmetry, made local by the choice of the propagator and the accompanying gauge fixing — see \cite{Lahoche_2026_background} for a detailed discussion. Matrix and tensor models, by contrast, require a dedicated treatment of these effects. Nevertheless, the structural similarity between the melonic sector of tensor models and the branched polymers of vector models suggests that the relational background method can be extended to higher-rank theories. An important difference, however, is that while vector and matrix models feature a single type of invariant interaction at quartic order, tensor models admit a wider variety of invariants, which may lead to multiple limiting theories.

From the non-perturbative point of view, the exact solvability of the LPA flow in the large-$N$ vector model also sheds light on our previous results for random matrices \cite{Lahoche_2026_background}. In the matrix case, the non-perturbative Wetterich equation does not reduce to a first-order equation integrable by characteristics, and both wave-function renormalization and multi-trace couplings enter at planar order. The vector calculation carried out here nevertheless indicates that the non-confining instability associated with double scaling should be looked for beyond polynomial vertex expansions around the origin, through a non-polynomial treatment of the effective potential in the non-confining sector. This non-perturbative investigation for matrices and tensors will be pursued in future work.

Finally, the method also raises several questions concerning the foundations of the RG itself and its relation to information geometry. Indeed, the background field selection procedure essentially amounts to replacing the initial distribution $P(\phi,\sigma)$, once the intermediate field $\sigma$ is introduced, with the conditional distribution $P(\phi\vert{}\sigma_0)$, where $\sigma_0$ is the saddle-point solution whose eigenvalues are frozen onto the Wigner distribution in the continuum limit. Let us emphasize that it is precisely this conditioning that gives substance to the notion of a relational scale. It is therefore interesting to examine the consequences of this conditioning at the level of the flow itself. The most general formulations of the RG, such as the Wegner--Morris formalism \cite{Wegner:1972ih,wegner1974some,morris1993exact}, imply the existence of fixed distributions, which are generally Gaussian \cite{cotler2023renormalization}. For the initial $O(N)$ model, the only allowed distributions have a propagator proportional to the identity and thus differ from one another only by global dilatations. According to Wegner in \cite{wegner1974some}, ordinary RG is itself defined only up to a dilatation, giving rise to what he refers to as redundant operators. Thus, global dilatations do not correspond to a genuine flow. By freezing a subset of the degrees of freedom, conditioning introduces a natural notion of scale and thereby allows for the definition of a genuine flow. An extended discussion of this subject can be found in \cite{Lahoche_2026_background}, and it will also be the focus of future investigations.

\section*{Acknowledgments}
V.L., for his part, extends his warm thanks to the swan of destiny and Isabelle Perreault.

\appendix 

\section{Critical exponent of the $p=2$ spin glass phase transition}\label{App1}
{
In this appendix, we recall the derivation of the critical exponent associated with the $p=2$ model at equilibrium, for which the reader is referred to \cite{de2006random,lahoche2023functional}. The first step consists in noting that the closed equation for the $2$-point function (the Schwinger--Dyson equation) reduces here to a closed equation for the mass, since no external momentum circulates in the effective loop:
\begin{equation}
M = m + u \int_0^4 \mathrm{d}p \, \frac{\rho(p)}{p + M}\,, \qquad \rho(p) = \frac{\sqrt{p(4-p)}}{2\pi}\,.
\end{equation}
The theory becomes critical when the effective mass $M$ vanishes, which corresponds to a critical temperature $m_c$ given by:
\begin{equation}
m_c = -u \int_0^4 \mathrm{d}p \, \frac{\rho(p)}{p}\,,
\end{equation}
so that:
\begin{equation}
h := m - m_c = M + u M \int_0^4 \mathrm{d}p \, \frac{\rho(p)}{p(p+M)}\,.
\end{equation}
For $M \ll 1$, the integral is dominated by the IR edge, where $\rho(p) \approx \pi^{-1}\sqrt{p}$, and therefore:
\begin{equation}
h \approx u \sqrt{M}\,,
\end{equation}
so that the value of the critical exponent associated with the mass scaling is indeed $0.5$.

\section{Burgers equation and shock}\label{App2}

Burgers' equation, initially introduced in the context of hydrodynamics, generically takes the following form, here without viscosity:
\begin{equation}
\frac{\partial u}{\partial t}+f(u) \frac{\partial u}{\partial x}=0\,.\label{eqBurger}
\end{equation}
The method of characteristics consists in writing $u(x(t),t)$ such that this function is stationary:
\begin{equation}
\frac{\dd}{\dd t} u(x(t),t)=0\,.
\end{equation}
Comparison with Eq.~\eqref{eqBurger} shows that:
\begin{equation}
\frac{\dd x}{\dd t}=f(u)\,.
\end{equation}
Here, $f(u)$ depends only on $u$, which is constant, and the equation integrates to give $x(t)=x_0+f(u) t$, so that, since $u(x(t),t)=u(x(0),0)\equiv u_0(x_0)$, we have:
\begin{equation}
u(x,t):=u_0(x-f(u(x(t),t)) t)\,.
\end{equation}
Differentiating with respect to $x$, we then find:
\begin{equation}
\frac{\partial u}{\partial x}= u_0^\prime(x-f(u_0(x_0)) t) \left(1-f^\prime (u_0(x_0)) t\frac{\partial u}{\partial x}\right)\,,
\end{equation}
which gives:
\begin{equation}
\frac{\partial u}{\partial x}=\frac{u_0^\prime(x_0)}{1+f^\prime (u_0(x_0))\,u_0^\prime(x_0)\, t}\,.
\end{equation}
The solution becomes singular when $t=t_c=-1/\big(f^\prime (u_0(x_0))\,u_0^\prime(x_0)\big)$, which can occur whenever $f^\prime(u_0(x_0))\,u_0^\prime(x_0)<0$, i.e. when the characteristic speed $f(u_0(x_0))$ decreases with $x_0$ (characteristics then cross). This is referred to as a shock, following the standard terminology in the literature, and the reader may consult Ref.~\cite{olver2014introduction} for instance.}

%\clearpage
\printbibliography[heading=bibintoc]

%\clearpage

\end{document}